# Validating an algebraic approach to characterizing resonator networks

Viva R. Horowitz [1] *, Brittany Carter [2,3,4], Uriel Hernandez [2,3,4], Trevor Scheuing [1], and Benjamín J. Alemán [2,3,4,5] *

[1]Physics Department, Hamilton College, Clinton, New York, 13323, United States
[2]Department of Physics, University of Oregon, Eugene, Oregon, 97403, United States
[3]Materials Science Institute, University of Oregon, Eugene, Oregon, 97403, United States
[4]Center for Optical, Molecular, and Quantum Science, University of Oregon, Eugene, Oregon, 97403, United States
[5]Phil and Penny Knight Campus for Accelerating Scientific Impact, University of Oregon, Eugene, Oregon, 97403, United States

* Corresponding authors: Viva R. Horowitz, vhorowit@hamilton.edu, Benjamín J. Alemán, baleman@uoregon.edu.

Resonator networks are ubiquitous in natural and engineered systems, such as solid-state materials, neural tissue, and electrical circuits. To understand and manipulate these networks—which are commonly modeled as systems of interacting mechanical resonators—it is essential to characterize their building blocks, which include the mechanical analogs of mass, elasticity, damping, and coupling of each resonator element. While these mechanical parameters are typically obtained from response spectra using least-squares fitting, this approach requires *a priori* knowledge of all parameters and is susceptible to large error due to convergence to local minima. Here we validate an alternative algebraic means to characterize resonator networks with no or minimal *a priori* knowledge. Our approach recasts the equations of motion of the network into a linear homogeneous algebraic equation and solves the equation with a set of discrete measured network response vectors. For validation, we employ our approach on noisy simulated data from a single resonator and a coupled resonator pair, and we characterize the accuracy of the recovered parameters using high-dimension factorial simulations. Generally, we find that the error is inversely proportional to the signal-to-noise ratio, that measurements at two frequencies are sufficient to recover all parameters, and that sampling near the resonant peaks is optimal. Our simple, powerful tool will enable future efforts to ascertain network properties and control resonator networks.

**Keywords**: resonator network, differential equation, numerical simulation, noisy system, singular value decomposition, inverse problem

# I. Introduction

Resonator networks are a ubiquitous [1] and diverse class of systems, found in both natural [2] and engineered contexts. They can range in scale from astronomical systems [3] to biological and neural networks [4–6], and from small-scale micromechanical lattices to large integrated circuits [7] and solid-state materials. These many-body systems exhibit rich collective behaviors, such as brain memory, quantum [8,9] and classical computation, and optical properties of solids, making them an important subject of study. To understand and engineer this behavior, it is necessary to characterize the network building blocks (e.g., neurons, micromechanical resonators, ions, etc.) and their connectivity. Despite many realizations, a useful and simple model for these networks is as a collection of coupled mechanical mass-spring resonators (**Fig. 1a**). Using this model, the building blocks are defined locally by the elasticity, mass, and damping of each resonator, while the connectivity is captured by coupling springs. For a linear response, the resonator network is governed by the equation of motion

$$M\ddot{\vec{x}} + B\dot{\vec{x}} + K\vec{x} = \vec{F} \tag{1}$$

where $M$, $B$, and $K$ are the mass, damping, and elasticity matrices, respectively, and $\vec{F}$ is the external force. Typically, the mechanical elements ($M, B, K$, and $\vec{F}$) of these systems are characterized by analyzing amplitude and phase spectra with non-linear least squares (NLLS), where the mechanical elements are fitting parameters. NLLS fitting requires both solving the coupled differential equations (i.e., determining a closed-form solution for $\vec{x}(\omega)$) and *a priori* knowledge in the form of initial guesses for each of the network's mechanical parameters. These initial guesses are crucial, since the NLLS solution may converge to a local minimum determined by the initial guesses, and poor initial guesses can lead to inaccurate estimates of the true state of the network [10]. Moreover, as the number of parameters increases, the optimization problem becomes more complex, and the number of potential local minima generally increases, making NLLS more inaccurate [10].

Recently, we developed an alternative, non-regressive algebraic approach to circumvent the issues of NLLS [10], a method we call Network Mapping and Analysis of Parameters, or NetMAP [11]. In this approach, we transform Eq. (1) into

$$\mathcal{M}(\omega)\vec{Z}(\omega) = \vec{f} \tag{2}$$

where $\mathcal{M}(\omega) = -\omega^2 M + i\omega B + K$, $\vec{f}$ is the amplitude of $\vec{F}$, and $\vec{Z}(\omega)$ is the complex response vector of the network which includes the amplitude and phase for each resonator in the network measured at a frequency $\omega$ (**Fig. 1b**). We then measure and construct $\vec{Z}(\omega)$ at two or more values of $\omega$, and use these measurements to rearrange Eq. (*2*) into an augmented homogeneous equation

$$\mathcal{Z}\vec{p} = \vec{0} \tag{3}$$

where $\mathcal{Z}$ is a matrix of known, measured quantities determined by $\vec{Z}(\omega)$. The vector $\vec{p}$ consists of the desired, unknown elements of $M, B, K$, and $\vec{f}$ (**Fig. 1b**, and see Methods), and the solution space for

$\vec{p}$ is the null-space of $\mathcal{Z}$. In contrast to NLLS, NetMAP does not require *a priori* knowledge (i.e. neither exact nor approximate knowledge of the elements of $M, B$, or $K$) nor iterative computation, but instead solves for $\vec{p}$ directly with as few as two response vector measurements ($\vec{Z}(\omega)$). In Ref. [11] we used NetMAP to determine $\vec{p}$ for small clusters of graphene NEMS resonators and we found excellent agreement with expected parameter values and broader spectral response.

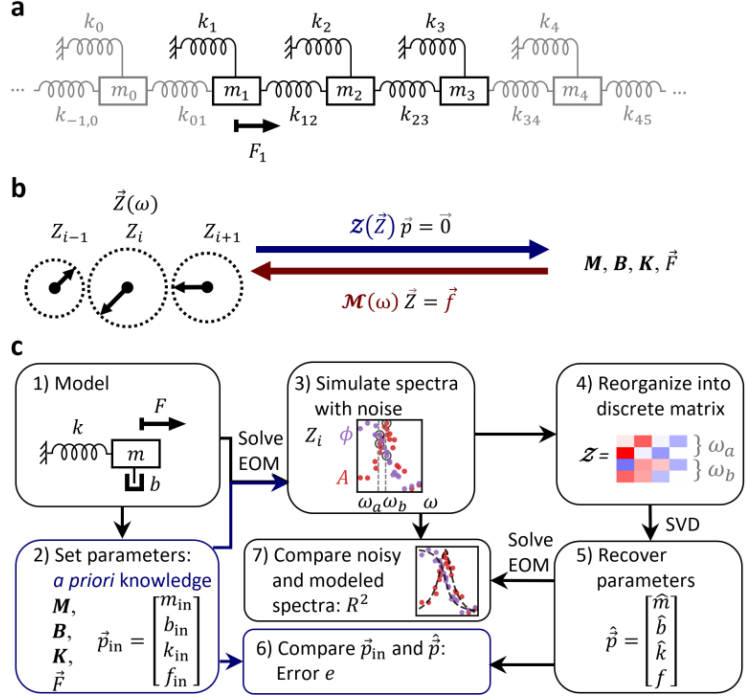

**Figure 1. a)** A resonator network as a chain of mass and spring oscillators. **b)** We calculate the relationship between the complex amplitude $\vec{Z}(\omega)$ of the masses and the underlying parameters. **c)** Steps of the validation process: 1) Modeling a mass-spring network, such as a monomer, shown here, or dimer. 2) Setting the parameters $\vec{p}_{in}$ to simulate. 3) Solving the equations of motion (EOM) and simulating spectra with noise. 4) Using simulated measurements of amplitude and phase at discrete frequency points to construct a matrix. 5) Using Singular Value Decomposition (SVD) to recover the parameters $\hat{\vec{p}}$, and scaling the result as needed. Here $f$ is shown without a hat to identify it as a known quantity while the others are scaled to $f$. 6) Calculating the percent error $e$ to compare the recovered parameters to the set parameters. 7) Calculating the expected spectra for the recovered parameters, and calculating $R^2$ to compare the simulated data and the expected curve. We find that $R^2$ is correlated with error $e$.

While NetMAP is a promising new technique, there are many unanswered questions regarding its accuracy in predicting the mechanical parameters of the network: How does the number $n$ of response vectors, their signal-to-noise ratio (SNR), or the frequency at which they were measured affect the accuracy of NetMAP? How do the actual values of the mechanical parameters or the dimension of the solution space of Eq. (3) affect the accuracy? To answer these fundamental questions, we simulate noisy response vectors for single-resonator (monomer) and coupled-pair (dimer) network clusters using predetermined mechanical parameters and then we use NetMAP to predict these parameters. We find that the accuracy of NetMAP predictions improves with the SNR, as does measuring response vectors near spectral peaks. While measuring additional vectors moderately improves accuracy, we find a minimum of two response vectors is sufficient, thus requiring a remarkably small number of measurements. Moreover, the dependence of the accuracy on the null-space dimension is nuanced; the accuracy of a 1D null-space solution varies with the input parameters.

## II. Methods

To assess the accuracy of NetMAP, we statistically compare the input parameters to the recovered parameters and the actual simulated spectra to the expected spectra from the model. The steps of the validation process are shown in **Fig. 1c** (see Supplemental Material[i] for details.) In the first step, we choose the size of mass and spring network to model ($N_\text{cluster}$). In this work, we modeled the monomer (a single resonator, $N_\text{cluster} = 1$) and the dimer (two coupled resonators, $N_\text{cluster} = 2$). We then set the network parameters $\vec{p}_\text{in}$ to fixed numerical values (step 2) and input them into the analytical solution of the equations of motion (EOM) to obtain an exact, noise-free complex response function for each resonator, $z_i(\omega)$ (see Supplemental Material). The response vector $\vec{z}(\omega)$ solves Eq. (2) exactly and has $N_\text{cluster}$ complex components.

To simulate a random experiment (step 3), we add real and imaginary noise to each response component to obtain $Z_i(\sigma, \omega) = z_i(\omega) + \Gamma_{x,i}(\sigma, \omega) + i\Gamma_{y,i}(\sigma, \omega)$, where $\Gamma(\sigma, \omega)$ is a pseudorandom number drawn from a zero-centered normal distribution of variance $\sigma^2$ and regenerated for each frequency $\omega$. Next (step 4), we select a set of noisy response vectors $\vec{Z}(\sigma, \omega)$ at $n$ discrete frequencies (*e.g.* $\omega_a$ and $\omega_b$ in **Fig. 1c**, so $n = 2$). Each $\vec{Z}(\omega)$ provides two vector equations corresponding to the real and imaginary parts of Eq. (2), for a total of $2nN_\text{cluster}$ linear equations, which we use to populate the matrix elements of $\mathcal{Z}$. The matrix $\mathcal{Z}$ has dimensions $2nN_\text{cluster} \times N$, where $N = \dim(\vec{p})$. (See Supplemental Material for general closed form of $\mathcal{Z}$ for the monomer and dimer.)

To obtain the predicted parameters vector $\hat{\vec{p}}$ (step 5), we use the simulated $\mathcal{Z}$ to solve $\mathcal{Z}\vec{p} = \vec{0}$ with a computational Singular Value Decomposition (SVD) algorithm. An algebraic approach is not standard but is straightforward for describing resonator systems. NetMAP uses singular value decomposition (SVD), an algebraic approach to solving systems of equations with applications in medical imaging [12], antenna arrays [13], planar transmission lines [14], and movie recommendations [15]. SVD is similar to eigensystem solvers, with singular values instead of eigenvalues and singular vectors instead of eigenvectors, while allowing the matrix $\mathcal{Z}$ to be rectangular rather than square. In using SVD to solve $\mathcal{Z}\vec{p} = \vec{0}$, we seek a singular value of zero and its corresponding singular vector, which is a solution for the physical parameters $\vec{p}$. The solution space for non-trivial $\hat{\vec{p}}$ will be at minimum one-dimensional (1D), but may have higher dimension. The solution-space dimension provided by SVD is open to interpretation, but commonly determined by the number of singular values with $\lambda \ll 1$. We explore the accuracy of NetMAP with 1D, 2D, and 3D solution spaces. When we define higher-dimension solution spaces, we use the smallest $\lambda$ and their corresponding singular vectors $\hat{\vec{p}}_\lambda$ in order of increasing value. In the 1D case, we scale $\hat{\vec{p}}$ so that the force $\hat{f}$ equals the input force $f$. For higher-dimensional solution spaces, we supply additional parameter constraints (see Supplemental Material).

Finally, to test the NetMAP parameter predictions, we replicate steps 3-5 of the simulation for a total of 1000 runs to generate a sample distribution for $\hat{\vec{p}}$, and then use a one-sample statistical $t$-test,

$$t_0 \equiv \frac{|(p_{in})_j - \hat{p}_j|}{s.e._{\cdot j}},$$

to quantify the agreement between the $j^{\text{th}}$ predicted network parameter $\hat{p}_j$ and the corresponding input parameter $p_{j,\text{in}}$. For each $\hat{\vec{p}}$ from the sample, we also compute the fractional error for each parameter $e_j = |\hat{p}_j - (p_{\text{in}})_j|/(p_{\text{in}})_j$, and we calculate the correlation coefficient ($R_i^2$) between the expected spectra $\hat{Z}_i(\omega)$ and the simulated noisy spectra $Z_i(\omega)$.

## III. Case studies

### A. A Monomer

To demonstrate NetMAP's algebraic approach, we first consider a test case with a lightly damped monomer. For a general monomer, the noisy spectrum is given by

$$Z(\sigma, \omega) = \frac{f}{-\omega^2 m + i\omega b + k} + \Gamma(\sigma, \omega) + i\Gamma(\sigma, \omega).$$

For this test case, we arbitrarily choose $\vec{p}_{\text{in}}$ with parameters: mass $m = 4$ kg, damping coefficient $b = 0.01$ N/(m/s), spring stiffness $k = 16$ N/m, and force amplitude $f = 1$ N. Moreover, we set the input error for this demonstration to $\sigma = 5 \times 10^{-3}$ m. **Fig. 2a** shows a simulated spectrum for the amplitude ($A$) and phase ($\phi$) of $Z(\sigma, \omega)$; **Fig. 2b** shows the real and imaginary parts of the same $Z(\sigma, \omega)$ plotted in the complex plane. To obtain $\hat{\vec{p}}$, we use $Z(\sigma, \omega)$ evaluated at the two frequencies $\omega_a$ and $\omega_b$ shown in **Fig. 2a,b**, which we then use to construct $\mathcal{Z}$. For a monomer sampled at two frequencies, the equation $\mathcal{Z}\vec{p} = \vec{0}$ written explicitly is:

$$\begin{pmatrix} -\omega_a^2 X(\omega_a) & -\omega_a Y(\omega_a) & X(\omega_a) & -1 \\ -\omega_a^2 Y(\omega_a) & \omega_a X(\omega_a) & Y(\omega_a) & 0 \\ -\omega_b^2 X(\omega_b) & -\omega_b Y(\omega_b) & X(\omega_b) & -1 \\ -\omega_b^2 Y(\omega_b) & \omega_b X(\omega_b) & Y(\omega_b) & 0 \end{pmatrix} \begin{pmatrix} m \\ b \\ k \\ f \end{pmatrix} = \begin{pmatrix} 0 \\ 0 \\ 0 \\ 0 \end{pmatrix} \quad (4)$$

where $X(\omega) = \text{Re}(Z(\omega))$ and $Y(\omega) = \text{Im}(Z(\omega))$. The simplest way to build $\mathcal{Z}$ is to read the values from complex plane plots, as in **Fig. 2b**. By solving Eq. (4) with SVD, we obtain solutions for the physical parameters $\hat{m}$, $\hat{b}$, and $\hat{k}$. The observed $t$-statistics for 1D, 2D, and 3D solution spaces is shown in Table 1. Given our large sample size (1000), the $t_0$ approximates the number of standard error intervals the predicted value is from the expected. All $p$-values exceed the $\alpha = 0.05$ by at least a factor of 10, so we conclude all NetMAP predicted parameters for this trial agree with the set input values $\vec{p}_{\text{in}}$.

**Table 1:** Observed $t$-statistics and associated $p$-values for network parameters for the monomer shown in **Fig. 2**. Results for the 1D, 2D, and 3D solution spaces are shown.

| $\hat{p}_j$ | $t_0 \equiv \dfrac{|(p_{\text{in}})_j - \hat{p}_j|}{s.e._j}$ | $p$-value |
|---|---|---|
| $\hat{m}_{1D}$ | 0.0803 | 0.9360 |
| $\hat{b}_{1D}$ | 0.2961 | 0.7672 |
| $\hat{k}_{1D}$ | 0.0805 | 0.9358 |
| $\hat{b}_{2D}$ | 0.2961 | 0.7672 |
| $\hat{k}_{2D}$ | 0.3719 | 0.7100 |
| $\hat{b}_{3D}$ | 0.3657 | 0.7146 |

A useful measure of the accuracy of NetMAP is the fractional discrepancy $\Delta p_j/(p_{\text{in}})_j = (\hat{p}_j - (p_{\text{in}})_j)/(p_{\text{in}})_j$. The $\Delta p_j/(p_{\text{in}})_j$ distributions for the 1D and 2D null-spaces are shown in **Fig. 2c**. The 1D result recovers all parameters within 0.04% of the input values in 95% of the trials (**Fig. 2c**). The 2D discrepancy results are similar or better than the 1D; for example, the 95% confidence range for the elasticity is $\sim 2.5 \times 10^{-5}$ %. The 95% confidence interval for the mean $\Delta p_j/(p_{\text{in}})_j$ are much tighter; for the 1D mass, $\dfrac{\Delta m_{1D}}{m_{\text{in}}} = (-4.9 \pm 6.8) \times 10^{-4}$ %.

As an additional test of NetMAP accuracy, we compare the predicted $\hat{Z}(\omega)$ to the noisy simulated $Z(\sigma,\omega)$ with correlation analysis. $\hat{Z}(\omega)$ is shown as a black-dashed curve in **Fig. 2a,b**. Using $n_R = 100$ simulated spectral data points, we compute the correlation coefficients $R_X^2$ and $R_Y^2$, where $X = \text{Re}(Z)$ and $Y = \text{Im}(Z)$. For the data in **Fig. 2b**, the deviation of the $R^2$ values from unity are $1 - R_X^2 = 8.6 \times 10^{-8}$ and $1 - R_Y^2 = 8.0 \times 10^{-8}$, or an average of $1 - R^2 = 8.3 \times 10^{-8}$. These $R^2$ values indicate that the NetMAP prediction accounts for essentially all the variation of simulated spectra, despite calculating $\hat{Z}(\omega)$ from response vectors at just $n = 2$ frequencies.

The fractional error, $e_j = |\hat{p}_j - (p_{\text{in}})_j|/(p_{\text{in}})_j$ is a useful metric to assess the accuracy of NetMAP. However, an experimentalist without knowledge of input parameters cannot calculate the fractional error but they can calculate the coefficient of determination $R^2$. To elucidate the relationship between the average $R^2$ and $e_j$, we plot the average error $\langle e \rangle = \dfrac{1}{N-D}\sum e_j$ versus $1 - R^2$ for each of the 1000 simulated trials and for the 1D and 2D null-spaces (**Fig. 2d**). For both dimensions, we observe a linear correlation ($R_{1D}^2 = 0.67, R_{2D}^2 = 0.4$). For the 1D case, we observe $\langle e \rangle$ decreases as $\sqrt{1 - R^2}$ ($R_{1D}^2 \sim 0.74, p \ll 0.001$). The correlation between $1 - R^2$ and error provides a means for an experimentalist to assess the accuracy of recovered values without *a priori* information. The corresponding distributions for error $\langle e \rangle$ are provided in **Fig. 2e.** While the 2D error is lower than the 1D ($\langle e \rangle_{1D} = 0.0144 \pm 0.0003\%$ vs. $\langle e \rangle_{2D} = 0.0039 \pm 0.0001\%$), we had to specify two input parameters $(p_{\text{in}})_j$ to solve for $\hat{\vec{p}}$. In general, for a D-dimensional null-space, $D$ number of input parameters $\vec{p}_{\text{in}}$ are required to obtain a solution, which is a disadvantage in terms of *a priori* knowledge. Generally, 1D solutions are preferable and sufficient to recover the parameters with low error.

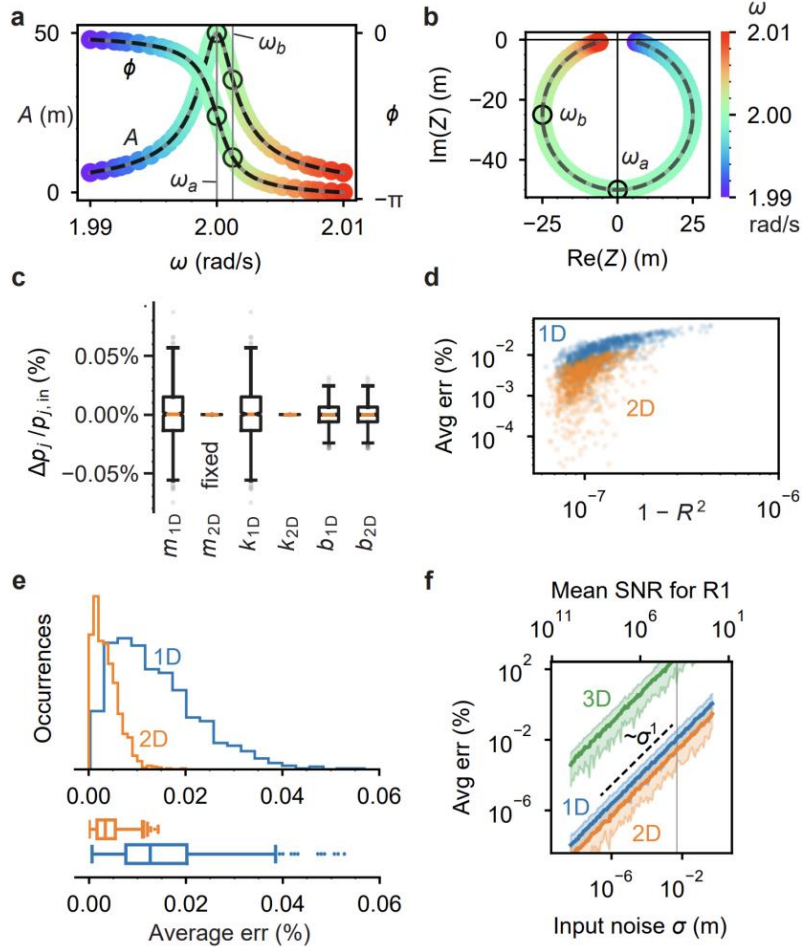

**Figure 2. a)** Amplitude $A(\omega)$ and phase $\phi(\omega)$ spectrum of the lightly damped monomer, with input parameters: mass $m = 4$ kg, damping coefficient $b = 0.01 \frac{N}{m/s}$, spring stiffness $k = 16$ N/m, and force amplitude $f = 1$ N. A subtle grey curve shows the exact spectrum while the datapoints in color show simulated spectrum measurements with standard deviation $\sigma = 0.005$ m. The black dashed line shows the output spectra $\hat{A}(\omega)$ and $\hat{\phi}(\omega)$. The color scale corresponds to (b). **b)** The complex amplitude $Z = Ae^{i\phi}$ is plotted in the imaginary plane, where the datapoints are phasors, $A$ is the distance from the origin, and phase $\phi$ is the polar angle. These are the same data as (a). Two measurements [$Z(\omega_a)$ and $Z(\omega_b)$, where $\omega_a = 2.0000$ rad/s and $\omega_b = 2.0013$ rad/s, black circles] are selected for input to NetMAP: one at the frequency of maximum amplitude and the other at a frequency corresponding to $\phi = -\frac{3}{4}\pi$. The remaining $Z(\omega)$ [colorful datapoints] are used only for validation, not for finding the recovered parameters. **c)** Box and whisker plot showing the spread in recovered parameters, $\Delta p_j/p_{j,\text{in}} = (\hat{p}_j - p_{j,\text{in}})/p_{j,\text{in}}$, for 1D and 2D solutions with 1000 runs of the same parameters. For 2D-SVD, additional a priori information is required: we fix $m = m_{in}$ in order to select the solution from the 2D solution space. **d)** Error $\langle e \rangle$ and $1 - R^2$ are correlated (1000 runs). We calculate $\langle e \rangle$ from a priori information, but we may predict it approximately from the $R^2$ value. **e)** Histogram and box plots of the percent error $\langle e \rangle$ (1000 runs), showing a half-normal distribution. **f)** Expanding to a range of input noise $\sigma$ for this lightly damped monomer (80 runs per $\sigma$), we find that the 2D-SVD solution is slightly more accurate than the 1D-SVD solution. The 3D solution is many orders of magnitude less accurate. The vertical grey line shows the input noise $\sigma = 0.005$ m appearing in other subfigures (a,b,c,d,e) of this figure. The average percent error and the standard deviation are related by a power law. The average curve is calculated as a mean of the logarithm of the average errors and the shaded regions indicate 95% of the runs.

So far, we have presented results for a fixed input noise ($\sigma = 5 \times 10^{-3}$ m.). To determine how the level of noise affects the error, we sweep $\sigma$ through several orders of magnitude and run the simulation 80 times per noise value. The error $\langle e \rangle$ vs. $\sigma$ is shown in **Fig. 2f**, with the $\sigma = 5 \times 10^{-3}$ m trial indicated with a vertical gray line. We also compute the signal-to-noise ratio defined as SNR $= \frac{A}{\sigma}$ and add it to the upper axis. We find that $\langle e \rangle$ for all solution space dimensions varies approximately linearly with $\sigma$ (*e.g.* for 1D, $\langle e \rangle = \beta \sigma^\alpha$ with $\alpha = 1.0004 \pm 0.0016$). The 2D-SVD error is the lowest, but is followed closely by the 1D error. The 3D error is several orders of magnitude larger than either 1D or 2D.

## B. A Dimer

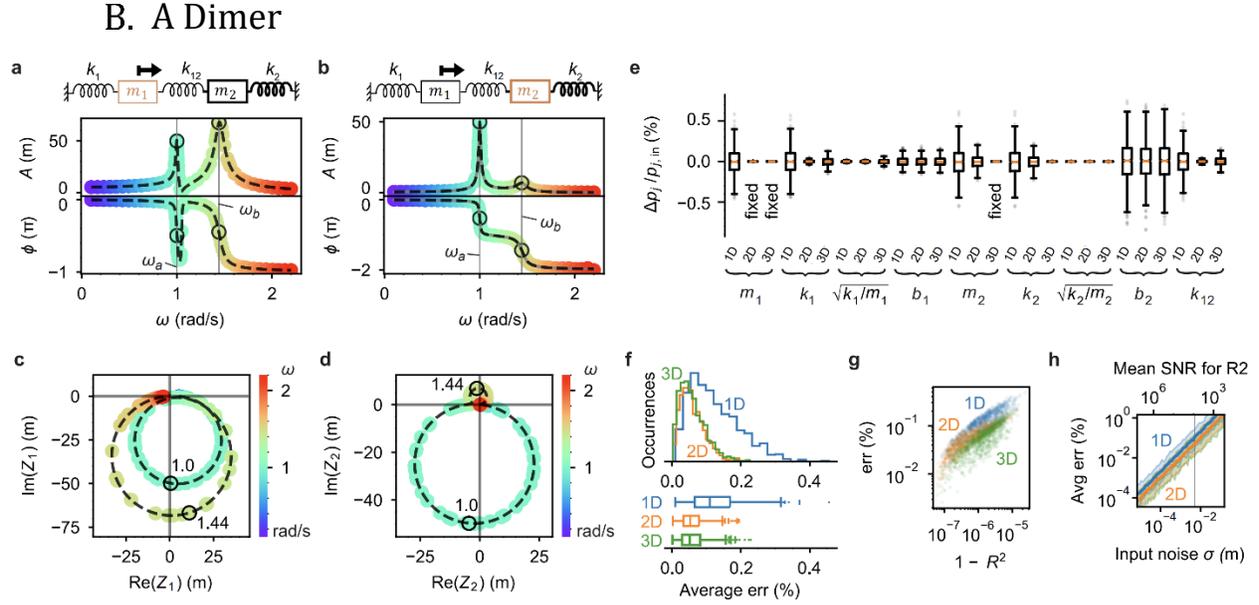

**Figure 3.** For an example dimer system with selected parameters, we demonstrate sampling at $n = 2$ frequencies, each at a resonance peak (black circles in (a,b,c,d)). **a)** Amplitude $A$ and phase $\phi$ spectra showing the motion of resonator 1 (R1). **b)** Spectra of resonator 2 (R2). **c)** Complex spectrum of R1, showing the same dataset as (a). **d)** Complex spectrum of R2, showing the same dataset as (b). **e)** Box and whisker plot showing the spread of recovered parameters as fractional discrepancy, $\Delta p_j / p_{j,\text{in}} = (\hat{p}_j - p_{j,\text{in}})/p_{j,\text{in}}$, for 1D, 2D, and 3D solutions over multiple trials. For 2D-SVD, an additional parameter, $m_1$, is fixed at $m_{1,in}$ in order to identify the solution within the 2D solution space. For 3D-SVD, $m_2$ is also fixed. **f)** Histogram and box plots of the average error $\langle e \rangle$. **g)** The average error $\langle e \rangle$ is correlated with $1 - R^2$. **h)** High SNR is key to minimizing the error. The error $\overline{\langle e \rangle}$ is proportional to the input noise $\sigma$.

We now similarly analyze a two-mass (dimer) resonator system (**Fig. 3**). For this system, we set the input simulation parameters as follows: $m_1 = 1$ kg, $m_2 = 10$ kg, $k_1 = 1$ N/m, $k_2 = 10$ N/m, coupling spring $k_{12} = 1$ N/m, $b_1 = b_2 = 0.1$ N/(m/s), $f_1 = 10$ N, and noise $\sigma = 0.005$ m. The force is applied to $m_1$. This case represents an anti-crossing scenario where the coupling splits the resonant frequency degeneracy. The noisy spectra of each resonator are shown in **Fig. 3a-d**, and we identify the two frequencies $\omega$ and corresponding responses (black circles) needed to construct $\mathcal{Z}$ and solve for $\hat{p}$. The fractional discrepancy $\Delta p_j / (p_{\text{in}})_j$ for 1D, 2D, and 3D solutions is shown in **Fig. 3e** (see Supple-mental Material for $t$-test results); we also include the discrepancy for the isolated resonance frequencies $\omega_j = \sqrt{k_j/m_j}$. We find that the damping of the undriven resonator, $b_2$, is the least accurate of the recovered parameters. Nonetheless, all parameters are recovered with an error standard deviation below 0.5%. The distributions for average fractional error $\langle e \rangle$ are provided in **Fig. 3f**, where $\langle e \rangle_{1D} = 0.123 \pm 0.002\%$, $\langle e \rangle_{2D} = 0.060 \pm 0.001\%$, and $\langle e \rangle_{3D} = 0.060 \pm 0.001\%$). Using $\hat{\vec{p}}$, we calculate and plot the recovered spectra $\hat{Z}_1(\omega)$ and $\hat{Z}_2(\omega)$, shown as black dashed lines in **Fig. 3a-d**. As before, we plot the average error $\langle e \rangle$ versus $1 - R^2$ (**Fig. 3g**) and the input noise $\sigma$ (**Fig. 3h**), and we observe a similar correlation with $1 - R^2$ and functional dependence $\langle e \rangle \propto \sigma$ for all null-space dimensions (see Supplemental Material).

## IV. Optimizing frequency selection

We have thus far demonstrated NetMAP on a monomer and a dimer system using the minimum number ($n = 2$) of response vector measurements. Even in the presence of noise, the 1D null-space solution recovers all network parameters with high accuracy. While any choice of at least two frequencies for response vector measurements will allow the analysis to proceed, we expect some frequencies to yield better results than others, and that sampling response vectors at more frequencies will yield different results. To test these ideas, we varied the number and value of the sampling frequencies for both a monomer and a dimer system while characterizing the accuracy of NetMAP.

### A. Monomer frequency optimization.

In order to assess the accuracy as a function of the number of measurement frequencies $n$, we first consider a moderately damped ($Q = 16$) monomer with $m = 4$ kg, $b = 0.4$ N/(m/s), $k = 10$ N/m, $f = 1$ N, $\sigma = 5 \times 10^{-3}$ m and we sweep $n$ from $n = 2$ to 25. We start with two points near the resonance frequency $\omega_{\text{res}} = 1.58$ rad/s and add additional points sequentially to the right (i.e. higher frequency) and left (lower frequency) of the peak, with $\Delta\omega = 0.01$ rad/s separation between measurement points (**Fig. 3a**, see Supplemental Material for the amplitude and phase spectra.). We replicate a given $\omega$ for a total of 100 runs. The SNR for the peak datapoint is $A/\sigma = 3164$. **Fig. 3b** shows the average error $\langle e \rangle$ vs. $n$ for 1D, 2D, and 3D null-space solutions, which we label as 1D-SVD, 2D-SVD, etc. The 1D error gradually decreases with $n$, but with diminishing returns as the number of measurement points becomes large; the 1D mean errors for $n \geq 7$ are statistically equivalent (see Supplemental Material for ANOVA post hoc analysis). The 2D error dependence on $n$ is similar the 1D, but plateaus after $n = 4$. However, while the 2D solution has lower error than the 1D for low $n$, the two solutions appear to converge for $n \geq 9$; at $n = 25$ the mean difference between 1D and 2D error is $0.008 \pm 0.002\%$. As with the first monomer test case, the 3D error is orders of magnitude larger than 1D or 2D error and does not decrease with $n$ but instead oscillates with $n$ as frequency points are incorporated to the left and right. Due to the high error of the 3D solution (sample distribution, $\langle e \rangle_{3D} = 8 \pm 15\%$), it is not recommended for this system.

To investigate the effect of frequency choice on the error, we consider two response vectors with frequencies $\omega_a$ and $\omega_b$ and measure $\overline{\langle e \rangle}$ for 1D and 2D null-spaces as we vary the frequencies across resonance. We plot $\overline{\langle e \rangle}$ vs. $(\omega_a, \omega_b)$ in **Fig. 4c**, and find that both the 1D and 2D solution are optimized near the peak resonance frequency ($\omega_{\text{res}} = 1.58$ rad/s). The "cross" patterns indicate that if either $\omega_a$ or $\omega_b$ is on resonance, then the other frequency can take nearly any value and the error remains small. For the 1D case, however, there is a narrow diagonal line of high error (68.3% ± 1.2% standard deviation), indicating that $\omega_a$ and $\omega_b$ must differ to obtain a low-error 1D solution. This diagonal line is absent in the 2D case, so that replicated $\vec{Z}(\omega)$ with the same frequency are possible and yield low error. To observe more general trends, we plot the average error for different choices of $\omega_a$ while taking an average of all $\omega_b$ values (**Fig. 4d**). This result shows that the optimal $\omega_a$ to extract $\vec{Z}(\omega_a)$ is near the resonant frequency, $\omega_{\text{res}}$. Moreover, by fixing $\omega_a = \omega_{\text{res}}$ we determine which $\omega_b$ will yield the lowest error. **Fig. 4e** shows the average error vs. the complex phase of the response vector $\vec{Z}(\omega_b)$ for varying $\omega_b$. From this plot, the error of the 1D solution reaches a minimum near $\phi_b = -3\pi/4$ and $\phi_b = -\pi/4$, and is greatest at $\phi_b = \phi_a \approx -\pi/2$. In accord with Fig. 4c, the error of the 2D solution is minimum near $\phi_b = \phi_a \approx -\pi/2$. We see similar optimal phase choices for other monomer examples (see Supplemental Material).

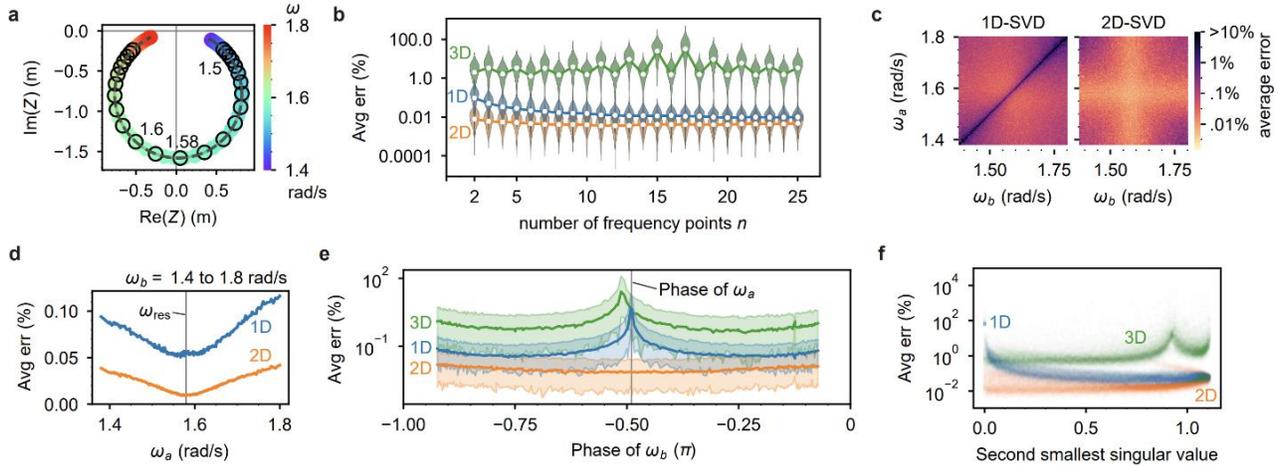

**Figure 4.** Comparing various choices of frequency measurements for analyzing a monomer with $m = 4$ kg, $b = 0.4\ \frac{\text{N}}{\text{m/s}}$, $k = 10$ N/m, $f = 1$ N, resonance frequency $\omega_{\text{res}} = 1.58$ rad/s, input noise $\sigma = 0.0005$ m. **a)** Input (circled datapoints) and output (dashed black curve) spectrum for a simulated monomer with 25 circled points used for analysis. **b)** In order to compare the number $n$ of frequency measurements, we plot the average error for m,k,b as a function of the number of frequency points used in the SVD analysis. The lines show the mean error $\overline{\langle e \rangle}$ of 100 simulated trials (shown as datapoints) for each number of frequency points. Violin plots show the distribution of $\langle e \rangle$. For this monomer system, the 2D-SVD solution (orange) has the least error, then the 1D-SVD (blue), and the 3D-SVD (green) has the largest error. The average percent error of the parameters $\vec{p}$ varies with the number of frequency points used in the SVD analysis. For this analysis, we add the frequency points in a specific order, starting with two points at each of the two amplitude peaks and sequentially adding points to the left and right of the peaks. The specific datapoints used for b are indicated in a. **cdef)** Comparing the choice of frequencies if there are $n = 2$ frequencies. **c)** The average error of the SVD solution varies with the choice of the two frequencies $\omega_a$ and $\omega_b$. The 1D solution fails when the two frequencies are the same. Both the 1D and 2D solution have better results when the measurements are taken near the peak of resonance. **d)** The average error varies with the measured frequency $\omega_a$. Here the average is taken over all the results shown in (c), with $\omega_b$ varying from 1.4 to 1.8 rad/s. **e)** For a double ($n = 2$ frequency) measurement of a monomer, the average error $\overline{\langle e \rangle}$ varies with the choice of frequencies measured. In this case, we fix $\omega_a$ at the resonance peak and sweep $\omega_b$. We find that the optimum second frequency occurs when the phase at $\omega_b$ is near $\phi_b = -\frac{\pi}{4}$ or $\phi_b = -\frac{3\pi}{4}$. The colored bands correspond to 95% of the solutions. **f)** The second smallest singular value $\lambda_2$ predicts the average error of the SVD solutions across all combinations of frequencies from trials shown in (c).

For the monomer under study, the 2D null-space solution generally has the lowest error. However, increasing the dimension of the solution space assumes the singular value of the additional degree of freedom is sufficiently small, i.e. $\lambda \ll 1$. If this value is not small, the lower dimension null-space should have lower error because the singular vector corresponding to the large singular value does not solve the homogeneous Eq (3). To test this idea, we plot the average error for 1D, 2D, and 3D null-spaces against the value of the second smallest singular value ($\lambda_2$) for each solution, as shown **Fig. 4f** (for this plot, we use the full set of swept frequency pairs from Fig. 4c.) We find that the 2D solution, which uses both the first and second singular vectors, is more accurate when $\lambda_2$ is small, while the 1D solution improves in accuracy as $\lambda_2$ grows. (See Supplemental Material for the 1D solution error as a function of both $\lambda_1$ and $\lambda_2$.) The 3D solution additionally uses the singular vector corresponding to the third singular value ($\lambda_3 \geq \lambda_2$). As expected, the 3D solution has consistently higher error than either 1D or 2D.

Our monomer case studies shed light on NetMAP best practices. For the above case study, the error generally decreases modestly as the number $n$ of measured response vectors increase; before plateauing, the 1D error decreases by $\sim 5\times$ with $n = 7$ and the 2D error decreases by $\sim 1.5\times$ with $n = 4$. In either case, $n = 2$ suffices. The 1D solution is preferred to the 2D because it requires minimal prior knowledge and has reasonably low error (i.e. $< 1\%$). However, the 2D error is consistently lower (mean of $\sim 8.3\times$ across all $n$) if $\lambda_1 \sim \lambda_2 \ll 1$. To obtain the lowest error with the 1D solution, $\vec{Z}(\omega)$ should be measured on resonance ($\phi_a \approx -\pi/2$) and $\pi/4$ radians off resonance, where $\phi_b = -3\pi/4, -\pi/4$. The sampling frequencies from Fig. 2 were selected in this way. Finally, given the error is proportional to the noise, it is beneficial to perform noise-filtered, long-integration experimental measurements of $Z(\omega_a)$ and $Z(\omega_b)$.

## B. Dimer frequency optimization

We now repeat the optimal-frequency analysis for a dimer system with the following input parameters: $m_1 = 8$ kg, $m_2 = 1$ kg, $k_1 = 2$ N/m, $k_2 = 7$ N/m, coupling spring $k_{12} = 5$ N/m, $b_1 = 0.5$ N/(m/s), $b_2 = 0.1$ N/(m/s), $f_1 = 1$ N, and noise $\sigma = 5 \times 10^{-5}$ m. Oscillating force is applied to $m_1$. The simulated spectra are shown in **Fig. 5a,b** (see Supplemental Material for corresponding amplitude and phase spectra); the spectral peaks inferred from the $Z_2(\omega)$ spectrum are at $\omega_{\text{res}} = 0.774$ rad/s and $\omega_{\text{res}} = 3.501$ rad/s. The higher frequency peak of $Z_2(\omega)$ has a markedly smaller maximum amplitude. The average errors for up to $n = 50$ sampled response vectors are shown in **Fig. 5c.** The error for 1D solutions starts at $\overline{\langle e \rangle}_{1D} = 0.50 \pm 0.30\%$ (standard deviation, SD) for $n = 2$, and improves with increasing $n$ down to $\overline{\langle e \rangle}_{1D} = 0.125 \pm 0.077\%$ (SD) at $n = 50$. Analyzing the 1D case with ANOVA (see Supplemental Material), we see that solutions with $n \geq 16$ have equivalent error and provide a modest ~2.5 × error improvement over the $n = 2$ solution. Compared to 1D solutions, the error for 2D solutions is higher, ranging from a minimum error of $2.37 \pm 1.88\%$ (SD) at $n = 12$ up to $215 \pm 208\%$ (SD) at $n = 2$. There is a pattern that repeats for every fourth additional frequency point: when a measurement is added near the stronger 0.774 rad/s peak, the 2D solution improves, and when a measurement is added near the 3.501 rad/s peak, the 2D solution becomes less accurate. Some of the measurements are numbered in **Fig. 5 a,b** to indicate the order in which they are included into **Fig. 5c**. The 3D solution has the overall lowest error for $2 \leq n \leq 11$, with a minimum $\overline{\langle e \rangle}_{3D} = 0.052 \pm 0.037\%$ (SD) at $n = 11$, but then suffers an abrupt loss in accuracy for $n \geq 12$ where the error jumps to $99 \pm 79\%$ (SD). To determine the optimal frequencies, we measure the average error with $n = 2$ and sweep both frequencies $\omega_a$ and $\omega_b$ across a range that covers both spectral peaks (**Fig. 5d-f**). Each solution space dimension has a high-error diagonal band corresponding to $\omega_a = \omega_b$. For the 1D solution, the lowest error occurs when one frequency is at the top of one peak and the other frequency is at the top of the other peak. The spectral peaks have the two highest SNR values that also sample both resonances. The 2D and 3D error patterns are more complex and more forgiving in terms of frequency choice; in both cases, it is sufficient to have one frequency near a spectral peak resonance, but the lowest error still occurs by sampling near each spectral peak.

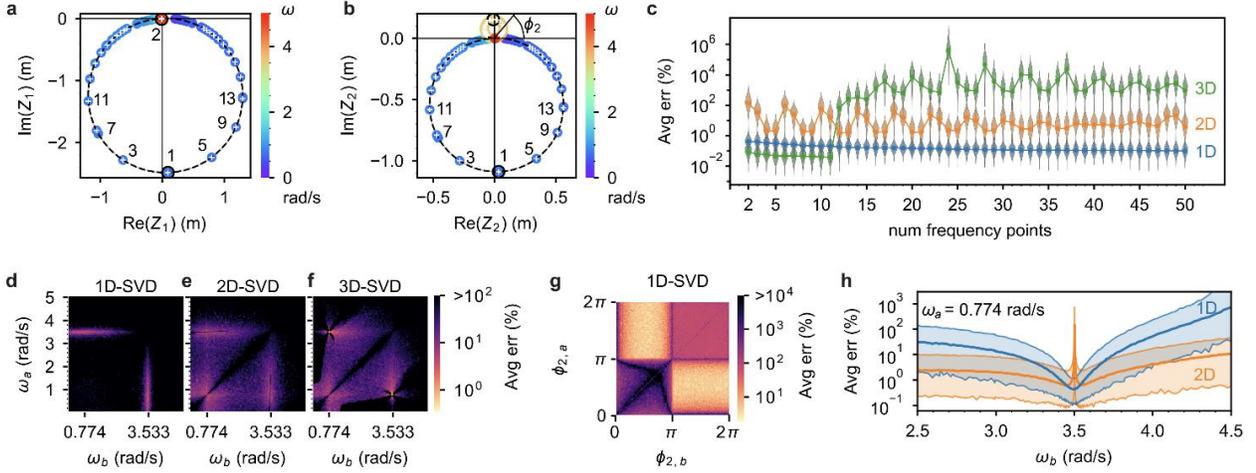

**Figure 5.** Analysis of optimal frequencies for NetMAP with an example two-mass (dimer) resonator system, with $m_{1,\text{in}} = 8$ kg, $m_{2,\text{in}} = 1$ kg, $k_{1,\text{in}} = 2 \frac{\text{N}}{\text{m}}$, $k_{12,\text{in}} = 5 \frac{\text{N}}{\text{m}}$, $k_{2,\text{in}} = 7 \frac{\text{N}}{\text{m}}$, $b_{1,\text{in}} = 0.5 \frac{\text{N}}{\text{m/s}}, b_{2,\text{in}} = 0.1 \frac{\text{N}}{\text{m/s}}$, $f_1 = 1$ N, and $\sigma = 5 \times 10^{-5}$ m. **a,b)** Complex spectrum of R1 (a) and R2 (b), showing the simulated spectrum with noise (colorful datapoints) and the spectra of the recovered parameters $\hat{Z}(\omega)$ (black dashed line, for 1 trial). White plus signs indicate datapoints that are added as $n$ increases from 2 to 50, and numbers indicate the order in which the points are included, with odd numbers for the larger peak and even numbers (not all shown) for the smaller peak illustrating how we alternate adding the frequency points. Only one resonant frequency, $\omega_{\text{res}} = 0.774$ rad/s (blue points), appears in the spectrum of R1 (a), but a small second peak at $\omega_{\text{res}} = 3.501$ rad/s (yellow points) appears in the spectrum of R2 (b). **c)** The average error varies with the number of datapoints $n$ used in the analysis, with 1D-SVD in blue, 2D-SVD in orange, and 3D-SVD in green. For this analysis, we add the frequency points in a specific order, starting with two points at each of the two amplitude peaks and sequentially adding points to the left and right of the peaks. Simulated with 99 trials per each of 49 values of $n$, totalling 4851 trials. **d)** If $n = 2$ datapoints are used, the 1D solution is most accurate when a measurement is taken at each of the two resonant peaks. **e,f)** The accuracy of the 2D and 3D solutions also varies with the pair of frequencies chosen. In general, it is less accurate for the two frequencies to be equal or nearly equal (dark diagonal lines). **g)** Replotting (d) in terms of R2 phase rather than frequency shows greater detail. Subfigures (defg) are simulated with 240000 trials. **h)** When $\omega_a$ is fixed at the lower resonant peak, $\omega_a = 0.774$ rad/s, then the accuracy of the 1D solution is optimized when $\omega_b$ is at the higher resonant peak, 3.533 rad/s (blue dip). The 2D solution is not accurate when each measurement is at one of the two resonant peaks (sharp orange peak). Shaded areas show 95% of simulated $\langle e \rangle$ results and lines show $\overline{\langle e \rangle}$. Simulated with 1600 trials per each of 199 values of $\omega_b$, totalling 318400 trials.

To see the detail of the 1D case, we plot the **Fig. 5d** dataset as a function of complex phases of $Z_2(\omega_a)$ and $Z_2(\omega_b)$ in **Fig. 5g** (see Supplemental Material for phase spectra plots). The $Z_2(\omega)$ phase $\phi_2(\omega)$ is one-to-one, unlike $\phi_1(\omega)$, so there is no ambiguity about which value of $\phi_2$ corresponds to which value of $\omega$. The phase plot shows consistently low error for quadrants II and IV, corresponding to recommended phase values $\phi_2 \in [0, \pi]$ for one measurement and $\phi_2 \in [\pi, 2\pi]$ for the other. In terms of the complex spectrum of $Z_2(\omega)$ (**Fig. 5b**), these phase values correspond to the upper and lower loops (spectral peaks). High error occurs in quadrants I and III, where the two response vectors are sampled on the same loop while the other is not sampled. Moreover, dark bands of high error occur when either phase is equal to 0 or $\pi$, which correspond to response vectors near the origin of the complex plane with near zero SNR. We further examine the behavior of the 1D and 2D solutions near the spectral peaks by fixing $\omega_a = 0.774$ rad/s and sweeping $\omega_b$, as shown in **Fig. 5h**. For the 1D case, error decreases as $\omega_b \to 3.533$ rad/s, which corresponds to the higher frequency spectral peak, where it reaches a minimum of 0.5%. The 2D case has similar behavior, but spikes to a maximum over 100% when $\omega_b = 3.533$ rad/s, indicating a failure in the 2D null-space accuracy when the response vectors are sampled directly on resonance. Thus, for a dimer with two resonant peaks and a 2D solution, measuring near—but not on—the spectral peaks is ideal.

## V. Accuracy varies with input parameters

So far, we have presented case studies of monomer and dimer resonator systems with fixed input parameters, $\vec{p}_{\text{in}}$. However, it is possible that the error behavior we observe depends on the choice of $\vec{p}_{\text{in}}$. To probe the dependence of the error on the input parameters themselves, we run a full, replicated $2^k$ factorial experiment for a general dimer system by varying the mechanical parameters (see Methods). We screened and ranked the mechanical parameters as predictors of average error for a dimer solved using 1D-SVD; including up to two-factor interactions, the order from most predictive to least is: $f$, $m_1$, $m_2$, $k_{12}$, and the $k_1 \cdot k_2$ interaction (see Supplemental Material for full effects model results). We previously discussed using $1 - R^2$ and $\lambda_2$ as indicators of average error (**Fig. 3g** and **Fig. 4d**).

To gain a deeper understanding of how the parameters $\vec{p}$ of the resonant system affect the SVD accuracy, we plot the average error as a function of varied parameters in a few example dimer cases (**Fig. 6**). Each cartoon in Fig. 6 shows one parameter in orange that we vary while keeping all others fixed. We identify the two peak frequencies $\omega_{\text{res}}$ using a peak-finding function [16] and plot them below each cartoon to illustrate their variation with the parameter. To improve the accuracy of the NetMAP results, we choose 4 additional frequencies on each side of the two peak frequencies $\omega_{res}$ for a total of $n = 10$ frequencies for analysis. To demonstrate the impact of signal to noise ratio (SNR) on the error as we vary each parameter, we plot the $\omega_{res}$ datapoints with higher SNR in orange, highlighting how the error falls as SNR increases. Increasing the driving force (**Fig. 6a**) or the non-driven mass $m_2$ (**Fig. 6b**) causes the SNR to increase and the error in the recovered parameters to decrease correspondingly. This matches our findings that error is inversely proportional to SNR, and we generally find that more accurate recovered parameters result for dimer systems with high force amplitude and larger $m_2$. We observe the error increases as we increase the coupling spring stiffness $k_{12}$ from 0 to 20 N/m (**Fig. 6c**), and this is usually true for dimers. For this dimer system, a lower $k_1$ corresponds to more accuracy (**Fig. 6d**), but this is not generalizable to all dimer systems. As $k_2$ is varied in **Fig. 6e**, the resonance peaks show an anticrossing near $k_2 = 20$ N/m, and the 1D solution loses accuracy near the anticrossing. This is a specific result for this particular dimer system. To see how an anticrossing affects a dimer system with different input parameters, we consider the system shown in **Fig. 6f**, with an anticrossing near $k_2 = 7$ N/m, and find, in contrast, that the 1D solution has higher accuracy near the anticrossing, suggesting that anticrossings may increase or decrease the error and the variation in error is likely associated with the SNR. Furthermore, to illustrate the effect on accuracy when the peak-finder fails to identify one of the resonance peaks, we allow the peak-finder to miss the higher $\omega_{\text{res}}$ peak for $k_2 = 10$ to 12 N/m, and find that the 1D solution loses accuracy while the 2D solution gains accuracy in that range. Thus, **Fig. 6** shows a detailed view of how the parameter values affect the accuracy of the algebraic solution.

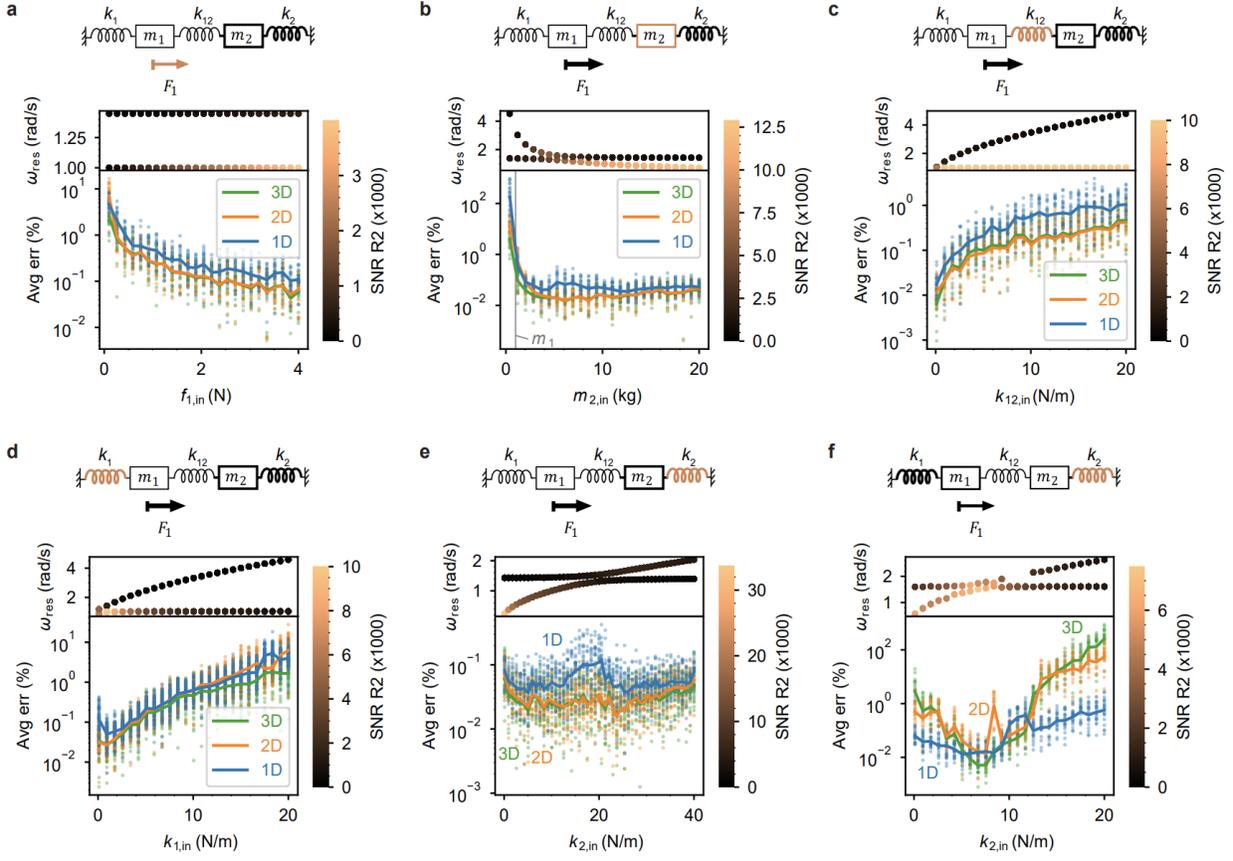

**Figure 6.** Varying dimer parameters affects the resonance frequencies and average error. In **(a - e)**, the input parameters are $m_1 = 1$ kg, $m_2 = 10$ kg, $k_1 = 1$ N/m, $k_{12} = 1$ N/m, $k_2 = 10$ N/m, damping $b_1 = b_2 = 0.1 \frac{N}{m/s}$, and force amplitude $f_1 = 10$ N, except the single parameter that is varied: force amplitude in (a), mass 2 in (b), coupling spring in (c), spring $k_1$ in (d), and spring $k_2$ in (e). In **(f)**, $k_2$ is varied and the input parameters are $m_1 = 5$ kg, $m_2 = 3$ kg, $k_1 = 12$ N/m, $k_{12} = 1$ N/m, $b_1 = 1 \frac{N}{m/s}$, $b_2 = 0.5 \frac{N}{m/s}$, and $f_1 = 10$ N. The resonator cartoons above each plot indicate the strength of parameters with line thickness, and the varied parameter is indicated in orange in the cartoon. The average error of the recovered parameters is plotted as a function of the varied parameter for three solution space dimensions, with 1D-SVD (blue), 2D-SVD (orange), and 3D-SVD (green). Error for individual simulations $\langle e \rangle$ is shown as datapoints and the average error across the simulations $\overline{\langle e \rangle}$ is shown as lines.

The frequency optimization and factorial studies provide broader NetMAP best practices for the dimer system. Driving with a higher amplitude force improves accuracy. For our case study in **Fig. 5**, the error decreases modestly as the number of frequency points increases, though $n = 2$ measured response vectors are sufficient. As it was for the monomer case, the 1D solution is preferred to the 2D or 3D because it requires minimal prior knowledge of the parameters, while still having reasonably low error, and because the 2D and 3D solutions have erratic accuracy, depending heavily on the particular frequency points sampled in our example in **Fig. 5**. To obtain the lowest error with the 1D solution, $\vec{Z}(\omega)$ should be measured at each of the two resonance peaks of the dimer spectrum. The two sampling frequencies in **Fig. 3** were selected in this way. As with the monomer case, the error is proportional to the noise, so it is beneficial to perform noise-filtered, long-integration experimental measurements of $Z(\omega_a)$ and $Z(\omega_b)$.

## VI. Discussion

For an experimentalist, NetMAP offers a means to analyze amplitude and phase data in order to reveal the physical parameters for each individual oscillator and the coupling between the oscillators. This phase-sensitive data is standard for lock-in amplifier measurements. However, for NetMAP to be useful, the experimentalist must measure the amplitude and phase for each oscillator in the network, which in some cases may be challenging. The SNR range we describe here (e.g. $10^3$, or 30 dB SNR) is attainable to experimentalists [11] working with a wide variety of systems, including electronic or micromechanical oscillator networks.

In cases where NetMAP returns an inadequate $R^2$ value, it may be advantageous to combine NetMAP and NLLS fitting by using the results from NetMAP as the initial guesses for an iterative solution, and thereby improving $R^2$. Since $R^2$ correlates with the error in the parameters, this is expected to improve the accuracy of the results.

We have shown how well NetMAP performs for a monomer (single mass) and dimer (double mass) system, but further investigation is needed to evaluate its performance in larger systems. Our calculations (see Supplemental Material) show that, even as the number of oscillators increases, only two discrete frequencies are needed for the required response vectors $\vec{Z}$. For a larger system, the size of the response vectors would scale with the number of oscillators, increasing the size of the matrix $\mathcal{Z}$, which may require solution spaces of higher dimension. Future work will explore NetMAP performance in larger systems.

# VII. Conclusions

In this study, we have presented and validated NetMAP, an efficient and reliable algebraic approach for calculating the physical parameters of resonator networks. We have tested NetMAP's accuracy for monomer and dimer systems as a function of the number of samples, choice of frequency points, and null-space dimension, and we provided a method for estimating the percent error using quantities available to an experimentalist. The spectra must be measured for each resonator at a minimum of two frequencies, ideally measured near resonance. We have developed NetMAP to measure the physical parameters of a MEMS system of coupled graphene resonators [11]. Future work includes exploring larger systems, including a two-dimensional grid of resonators; relating to other physical oscillators, including RLC circuits; simulating nonlinear springs; and considering systems where the topology of the system is unknown. NetMAP enables us to characterize the building blocks and connectivity of a diverse array of resonator networks. This offers to enhance our ability to design, tune, and program engineered resonator networks, such as micromechanical systems, and to better understand natural resonator systems, such as neural networks.

## Author Contributions

## Acknowledgement

We gratefully acknowledge conversations with Eric Corwin, James Sartor, Pieter Vandenberg, and Çağlar Girit. This work was supported by Hamilton College, a University of Oregon Reneé James grant, and NSF grant # CMMI-2128671.

# Supplemental Material: Validating an algebraic approach to characterizing resonator networks

Viva R. Horowitz [1] *, Brittany Carter [2,3,4], Uriel Hernandez [2,3,4], Trevor Scheuing [1], and Benjamín J. Alemán [2,3,4,5] *

[1]Physics Department, Hamilton College, Clinton, New York, 13323, United States
[2]Department of Physics, University of Oregon, Eugene, Oregon, 97403, United States
[3]Materials Science Institute, University of Oregon, Eugene, Oregon, 97403, United States
[4]Center for Optical, Molecular, and Quantum Science, University of Oregon, Eugene, Oregon, 97403, United States
[5]Phil and Penny Knight Campus for Accelerating Scientific Impact, University of Oregon, Eugene, Oregon, 97403, United States
* Corresponding authors: Viva R. Horowitz, vhorowit@hamilton.edu, Benjamín J. Alemán, baleman @uoregon.edu.

## Contents

# 1. Symbol list

## 1.1 Matrices

$\mathcal{M}$ matrix of desired unknown values

$\mathcal{Z}$ matrix constructed from measured or noisy simulated values of the spectrum at discrete frequency points

**M** mass matrix

**B** damping matrix

**K** stiffness or elasticity matrix

$D$ = nullity of $\mathcal{Z}$, or solution space dimension, where the solution space is the null-space of $\mathcal{Z}$. The value of $D$ is open to interpretation because $\mathcal{Z}$ has noise.

## 1.2 Vectors of complex spectra

$\vec{Z}(\omega)$: measurement vector or response vector (with noise), with elements $Z_1(\omega), Z_2(\omega), \ldots Z_{N_{\text{cluster}}}(\omega)$ for each resonator $i$ in the cluster.

$\vec{z}(\omega)$: simulated response vector (without noise) with elements $z_1(\omega), z_2(\omega), \ldots z_{N_{\text{cluster}}}(\omega)$ for each resonator in the cluster.

$\Gamma_{x,i}(\sigma, \omega)$: A pseudorandom number generated from a Gaussian distribution (standard deviation $\sigma$) to add noise to the real part of $Z_i$ for the $i^{\text{th}}$ resonator at a given discrete frequency $\omega$.

$\Gamma_{y,i}(\sigma, \omega)$: A pseudorandom number generated from a Gaussian distribution (standard deviation $\sigma$) to add noise to the imaginary part of $Z_i$ for the $i^{\text{th}}$ resonator at a given discrete frequency $\omega$.

$\omega$ angular frequency. The resonator network response frequency is assumed to be equal to the driving frequency for the steady state solution.

$\omega/2\pi$ frequency in MHz

$\vec{X}(\omega)$: The real part of $\vec{Z}(\omega)$, that is $\vec{Z}(\omega) = \vec{X}(\omega) + i\vec{Y}(\omega)$

$\vec{Y}(\omega)$: The imaginary part of $\vec{Z}(\omega)$

$\vec{A}(\omega)$: The amplitude of $\vec{Z}(\omega)$, that is $Z_i(\omega) = A_i e^{i\phi_i}$,

$\vec{\phi}(\omega)$: The phase of $\vec{Z}(\omega)$

## 1.3 Counts, indices, and more

$N_{\text{unknowns}}$: total number of unknown parameters

$N_{\text{cluster}}$: Number of oscillating masses. $N_{\text{cluster}} = 1$ for monomer, 2 for dimer

$n$: number of frequencies selected for analysis. If $n = 2$, we call it a double measurement.

$\omega_a, \omega_b$ or $\omega_1, \omega_2, \ldots, \omega_n$: input frequencies, the set of frequencies selected for analysis.

$n_t$: number of trials (number of repeats + 1)

$i$ = each resonator, from 1 to $N_{\text{cluster}}$ (e.g. $Z_i$, $\phi_i$, $R^2_{A,i}$, $R^2_{\phi,i}$)

i = imaginary constant

$\vec{p}$: parameters vector. It includes all unknown parameters and possibly 1 or more known parameters for scaling the solution. The order of the parameters in the parameters vector must be consistent with the order of the columns of $\mathcal{Z}$.

$\vec{p}_{\text{in}}$: input parameters. This is the *a priori* information.

$\hat{\vec{p}}$: the parameters output by NetMAP (Network Mapping and Analysis of Parameters). Components of this vector are $\hat{m}_1, \hat{b}_1, \hat{k}_1$, et cetera.

$N = \dim(\vec{p})$ the number of parameters

$j$: index of each parameter from 1 to $N$, e.g. parameter $p_j$, error $e_j$. In Fig. 1a, $p_1 = m, p_2 = b, p_3 = k$.

$\hat{\vec{Z}}, \hat{\vec{A}}, \hat{\vec{\phi}}, \hat{\vec{X}}, \hat{\vec{Y}}$: The complex spectrum vector and related real-valued parts, calculated from $\hat{\vec{p}}$.

$r$: number of trials for a factorial experiment

R1: resonator 1. The resonator we are driving with an oscillating force.

R2: resonator 2 in a two-mass (dimer) system. The resonator we are not directly driving.

$n_R$: number of frequencies for calculating $R$-value

$R^2_{A_i}$: coefficient of determination comparing the amplitude spectrum $\hat{A}_i(\omega)$ to the measured amplitude $A_i(\omega)$.
$R^2_\phi$: coefficient of determination comparing $\hat{\phi}_i(\omega)$ to $\phi_i(\omega)$
$R^2_{X_i}$: coefficient of determination comparing $\hat{X}_i(\omega)$ to $X_i(\omega)$
$R^2_{Y_i}$: coefficient of determination comparing $\hat{Y}_i(\omega)$ to $Y_i(\omega)$
$R^2$: The average of $R^2_{X_i}$ and $R^2_{Y_i}$, taking the average over all resonators $i$

$k$: monomer spring constant

$k$: number of parameters in factorial. Used in the context: $2^k$.

## 1.4 Force

$\vec{F}$: force vector, with components for each oscillating mass. The capital notation indicates that it is time-dependent (oscillating). $\dim(\vec{F}) = N_{\text{cluster}}$

$F_1$ or $F$: the oscillating force pushing mass 1.

$\vec{f}$, force vector, just the amplitude (with the time-dependent part divided out). $\vec{F} = \vec{f} \cos \omega t$

$f_1$ or f, the force amplitude pushing mass 1. $F_1 = f_1 \cos \omega t$

## 1.5 Results of singular value decomposition

$\lambda_1, \lambda_2, \ldots, \lambda_N$ the singular values of $\mathcal{Z}$, sorted from smallest to largest
$\vec{p}_1, \vec{p}_2, \ldots \vec{p}_N$, the associated singular vectors
$\lambda_1, \lambda_2, \ldots, \lambda_D$ the subset of singular values that correspond to the solution space
$\vec{p}_{1,\text{un}}$ the unscaled singular vector associated with the smallest singular value $\lambda_1$. It is normalized to length 1.
$\vec{p}_{2,\text{un}}$ the unscaled singular vector associated with the second smallest singular value $\lambda_2$. It is normalized to length 1.
$\hat{\vec{p}}_{1D}, \hat{\vec{p}}_{2D}, \hat{\vec{p}}_{3D}$: the recovered parameters, calculated using $D = 1, D = 2$, and $D = 3$, respectively.

## 1.6 Error
The error is a figure of merit to show how accurate recovered parameters are.

$$e_j = \frac{|\hat{p}_j - p_{j_{\text{in}}}|}{p_{j_{\text{in}}}} \cdot 100\% \qquad \text{For example, } e_m = \frac{|\hat{m} - m_{\text{in}}|}{m_{\text{in}}} \cdot 100\% \text{ is the percent error in mass.}$$

The discrepancy, $\Delta p = \hat{p}_j - p_{j,\text{in}}$. For example, $\Delta b = \hat{b} - b_{\text{in}}$.

$\bar{e}_j$: the logarithmic average of the error, taken over many trials, where $j$ indexes the parameter (for box and whisker plots)

$\langle e \rangle = \frac{e_1 + e_2 + \cdots + e_N}{N - D}$ average across parameters for an individual simulation trial

$\langle \bar{e} \rangle$: logarithmic average across trials first, then arithmetic average across parameters

$\overline{\langle e \rangle}$: arithmetic average across parameters first, then logarithmic average across trials.

# 2. Extended description of NetMAP
## 2.1 Resonator network theory

A system of equations for a resonator network can be organized into the matrix form:

$$M\ddot{\vec{x}} + B\dot{\vec{x}} + K\vec{x} = \vec{F}, \tag{S1}$$

where $\vec{x}$ is the displacements for each resonator, $\vec{F}$ is the oscillating forces driving each resonator, and $M$, $B$, and $K$ are matrices with all information about the inertia, damping, and elasticity of the network. If the force vectors are all oscillating with driving frequency $\omega$, then we guess an oscillating steady-state solution $\vec{x} = \vec{Z}e^{i\omega t}$ where $\vec{Z}$ is a constant complex amplitude vector with components $Z_i = A_i e^{i\phi_i}$, and obtain

$$-\omega^2 M\vec{Z} + i\omega B\vec{Z} + K\vec{Z} = \vec{f}, \tag{S2}$$

where $\vec{f}$ is the amplitude of the force vector, $\vec{F} = \vec{f}e^{i\omega t}$. More compactly, the equation of motion system can be written

$$\mathcal{M}(\omega)\vec{Z}(\omega) = \vec{f} \tag{S3}$$

where the symmetric matrix $\mathcal{M}(\omega) = -\omega^2 M + i\omega B + K$ contains complete information about the inertia, elasticity, and damping of the network.

## 2.2 Solving the equations of motion

In order to simulate spectra to validate NetMAP, our approach is to set the input values $\vec{p}_{\text{in}}$, then calculate the spectra $Z(\omega)$ with added random Gaussian noise, thereby simulating an experiment (steps 2 and 3 of Fig 1c). To calculate the input spectra, we implement Cramer's rule in SymPy [1] to symbolically solve Eq. (S2) given the input values $\vec{p}_{\text{in}}$. This yields the expected response vector

$\vec{z}(\omega)$, where each vector component $z_i$ corresponds to the $i$th resonator. To simulate noisy measurements, we add noise from a random Gaussian distribution to each vector component:

$$Z_i(\omega) = z_i(\omega) + \Gamma_{x,i}(\sigma,\omega) + i\Gamma_{y,i}(\sigma,\omega). \tag{S4}$$

We add noise to the real and imaginary parts of $z_i$ rather than to the amplitude and phase because doing so better matches experimental results, especially at low amplitudes, where the phase becomes highly uncertain. We also use the same solution to the equations of motion to calculate the output spectra $\hat{Z}_i(\omega)$ for step 7 of Fig 1c, but we use the output parameters $\hat{\vec{p}}$ and do not add simulated noise.

While the method we present here would be appropriate for any resonator system, we consider as a particular case a micromechanical system where we measure the amplitude and phase of each resonant mass using scanning interference measurements (SIM). [2,3]

## 2.3 Determining the number of frequencies required for NetMAP

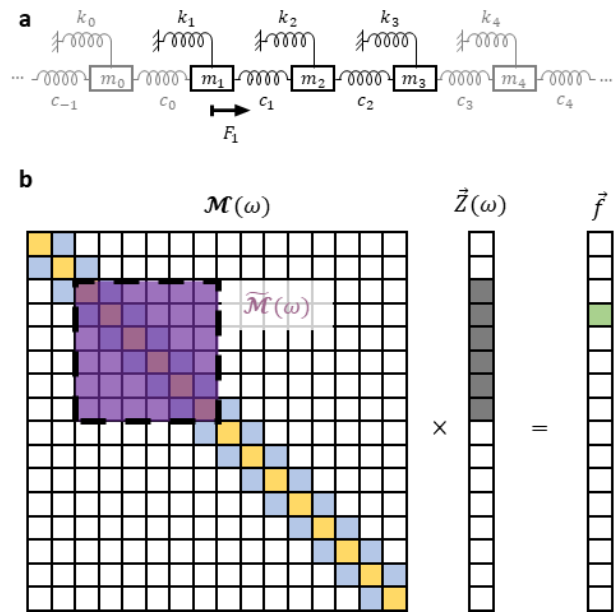

Figure S1 a) An oscillator network made up of a linear chain of masses and springs. b) Pictorial visualization of equation of motion showing the cluster response vector $\vec{Z}(\omega)$ and submatrix $\widetilde{M}(\omega)$.

1. Clusters

Consider the case of a linear chain of resonators (**Figure S1**a) where we apply an oscillating force of amplitude $f$ to one resonator in the chain. For a linear chain of resonators (**Figure S1**a), with coupling springs of stiffness $c_i$ only between nearest neighbors, the matrix $\mathcal{M}$ is tridiagonal (**Figure S1**b, left), with diagonal elements (yellow squares) of the form $-m_i\omega^2 + ib_i\omega + c_{i-1} + c_i + k_i$ and off-diagonal elements (blue squares) of the form $-c_i$, corresponding to Eq. (S2).

In an experimental system that is driven at one resonator (**Figure S1**b, right, light green square), a finite number of resonators will respond, indicated by the shaded squares for $\vec{Z}(\omega)$, which selects a relevant submatrix $\widetilde{\mathcal{M}}(\omega)$ (**Figure S1**b, purple square). Our response vector approach determines elements of $\widetilde{\mathcal{M}}(\omega)$ from the observations of $\vec{Z}(\omega)$ using Eq. (S3). However, the number of unknown elements in $\widetilde{\mathcal{M}}(\omega)$ and $f$ is larger than the number of equations provided by a measurement of $\vec{Z}(\omega)$ at one frequency, thereby requiring multiple measurements of $\vec{Z}(\omega)$ at distinct values of $\omega$. The number of required frequencies is determined by the number of unknowns in the network or network cluster, $N_{\text{unknowns}}$. Assuming a unique mass and damping for each resonator and setting the edge spring constants to zero, we write $N_{\text{unknowns}}$ in terms of the number of resonators in the network cluster:

$$N_{\text{unknowns}} = 4 \times N_{\text{cluster}}.$$

The full set of unknown parameters include $m_i$, $b_i$, $k_i$, all relevant $c_i$'s, and the amplitude $f$. The cluster size ($N_{\text{cluster}}$) is determined by the number of resonators with detectable oscillation, and will always be equal to or less than the total number of resonators in the network. Moreover, the response vector $\vec{Z}(\omega)$ will have $N_{\text{cluster}}$ non-zero components (although some might be zero or undetectable when sampled at some frequencies) and $\widetilde{\mathcal{M}}(\omega)$ will be a symmetric, tridiagonal $N_{\text{cluster}} \times N_{\text{cluster}}$ matrix. Each measurement of $\vec{Z}(\omega)$ at a fixed value of $\omega$ solves

$$\widetilde{\mathcal{M}}(\omega)\vec{Z}(\omega) = \vec{f} \tag{S5}$$

and thus provides two equations per resonator (due to real and imaginary parts):

$$N_{\text{equations}} = 2 \times N_{\text{cluster}}. \tag{S6}$$

Therefore, to determine the unknowns, we only need to acquire response vectors at a minimum of two frequencies—$\vec{Z}(\omega_a)$ and $\vec{Z}(\omega_b)$. These response vectors then provide a system of $4 \times N_{\text{cluster}}$ linear equations.

In cases where the size of the resonator network is finite, we may have the cluster encompassing the entire network, such that $N_{\text{cluster}} = N_{\text{unknowns}}$. In some experimental cases, motion of some resonators at a given $\omega$ will be small and undetectable by the apparatus and require acquisition of additional response vectors $\vec{Z}(\omega)$. When a resonator's motion is undetectable at some or all driving frequencies, then the number of linear equations is reduced; if the undetectable resonator is in the interior of the cluster, we lose six equations, while if at the edge of the cluster we lose four equations. Therefore, in these cases it is important to ensure that all resonators have a measurable amplitude at the two driving frequencies, or else to measure $\vec{Z}$ at additional driving frequencies.

### 2.4 Reorganizing the system of linear equations.

The response vectors at $n = 2$ frequencies, $\vec{Z}(\omega_a)$ and $\vec{Z}(\omega_b)$, together yield a system of $4 \times N_{\text{unknowns}}$ linear equations composed of

$$\widetilde{\mathcal{M}}(\omega_a)\vec{Z}(\omega_a) = \vec{f} \text{ and } \widetilde{\mathcal{M}}(\omega_b)\vec{Z}(\omega_b) = \vec{f} \tag{S7}$$

where each component $Z_i(\omega)$ has real amplitude $A_i(\omega)$ and phase $\phi_i(\omega)$:

$$Z_i(\omega) = A_i(\omega)e^{i\omega\phi_i}. \tag{S8}$$

We combine and reorganize Equations (S7) into the following single linear homogenous equation

$$\mathcal{Z}\vec{p} = \vec{0} \tag{S9}$$

where $\mathcal{Z}$ is a real-valued matrix with matrix elements that depend on the known measured quantities $A_i(\omega)$, $\phi_i(\omega)$, and $\omega$ (step 4 of Fig 1c). See below for the form of $\mathcal{Z}$ for the monomer and dimer case. When constructing $\mathcal{Z}$ in practice and calculating the error of $\vec{Z}(\omega)$, the vector $\vec{p}$—which we call the *parameters vector*—is an $N_{\text{unknowns}}$-dimensional vector composed of the unknown mechanical parameters of the cluster: $m_i$, $b_i$, $k_i$, all relevant $c_i$'s, and the force amplitude vector $\vec{f}$. Including $\vec{Z}(\omega)$ at additional driving frequencies will expand the dimensionality of $\mathcal{Z}$ and $\vec{p}$, which in general will over-determine the system of equations. Finally, we use singular value decomposition (SVD) to find the solution space of $\mathcal{Z}$, which determines the output parameters vector $\hat{\vec{p}}$.

Here, we simulate cases where the force amplitude vector is $\vec{f} = \langle f_1, 0 \rangle$, only driving the first resonator, but other force amplitudes $\langle f_1, f_2 \rangle$ are possible. Sampling response vectors from at least $n = 2$ drive frequencies, we then use equations of the form Eq. (S7) to construct $\mathcal{Z}$ and the system of linear equations $\mathcal{Z}\vec{p} = \vec{0}$. We apply a NumPy SVD solver [4,5] in Python to factorize and identify the solution space of $\mathcal{Z}$, and thereby generate a solution to recover the parameters $\hat{\vec{p}}$, where the hat symbol indicates that these are values recovered by the SVD analysis, not simulated measurements (step 5 of Fig 1c).

We obtain from the NumPy SVD solver the singular values $\lambda_1, \lambda_2, \ldots, \lambda_N$, sorted from smallest to largest, each associated with a singular vector $\vec{p}_1, \vec{p}_2, \ldots \vec{p}_N$, where $N = \dim(\vec{p}_i)$ is the number of parameters. With noise, the null value of $\mathcal{Z}\vec{p} = \vec{0}$ is not precisely zero, so we assume the singular vector associated with the smallest singular value $\lambda_1$ corresponds to the solution space. We may also consider multiple small singular values $\lambda_1, \lambda_2, \ldots, \lambda_D$ to be degenerate, even if they are not precisely equal to zero, and, in that case, we must find the solution within a $D$-dimensional solution space, for some positive integer $D \leq N$.

## 2.5 Scaling the singular vector: the question of solution space dimension

Since NumPy SVD solver provides normalized singular vectors, each of length 1, we must therefore scale the SVD output to identify the solution within the $D$-dimensional solution space and recover the numeric parameters. If we assume the solution space is 1D, we use $D = 1$ known values to scale the output vector. Here, we will require the force amplitude $f$ to equal the input value $f_{\text{in}}$, and thus we obtain a unique scaled solution $\hat{\vec{p}}_{1D} = \alpha \vec{p}_{1,\text{un}}$, where the normalized (i.e. unscaled) singular vector $\vec{p}_{1,\text{un}}$ corresponds to the smallest singular value $s_1$ and the scaling coefficient $\alpha$ allows $f$ to equal the input value. Here we use an input parameter, whereas in our experimental work, [2] we estimate a parameter.

The smallest singular value $\lambda_1$ will only equal zero exactly in the limit that the noise $\sigma$ approaches zero, and $\lambda_2$ is larger. Nevertheless, $\lambda_2$ may approximate zero, which would correspond to a 2D solution space. If we interpret the solution space as 2D, then $D = 2$ known values is sufficient to define a unique solution $\hat{\vec{p}}_{2D} = \alpha \vec{p}_{1,\text{un}} + \beta \vec{p}_{2,\text{un}}$ from within the 2D space, and, while any two choices of parameters are options, here we will fix both $f$ and $m_1$ to equal the input values and allow the other parameters to be scaled accordingly.

If we assume the solution space is 3D, we require $D = 3$ known values to define a unique solution $\hat{\vec{p}}_{3D} = \alpha \vec{p}_{1,\text{un}} + \beta \vec{p}_{2,\text{un}} + \gamma \vec{p}_{3,\text{un}}$; for these three known values, here we use $f, m_1$, and $m_2$ for the dimer or $f, m$, and $k$ for the monomer, leaving the damping $b$ as the only unknown parameter in the monomer case.

Ultimately, we obtain three possible solutions for the parameters, $\hat{\vec{p}}_{1D}, \hat{\vec{p}}_{2D}$, or $\hat{\vec{p}}_{3D}$, where the 1D solution is the easiest to obtain, requiring the least additional information. For an experimentalist, the choice of solution space dimension (the nullity $D$) may be informed by running simulations and by considering the size of the singular values. If the second smallest singular value $\lambda_2$ is low, then the 1D solution space may be insufficient and a 2D solution space may be more accurate, as shown in Fig 4d. In such a case, an experimentalist seeking to calculate the physical parameters using NetMAP has three choices: (1) assume a 2D solution space (or higher dimensional) and find a solution in that space, (2) use the 1D solution space solution, recognizing that the error will be larger, or (3) take additional spectral data to improve the accuracy of the 1D solution. In any of these situations, simulations may inform the choice.

In order to validate the accuracy of NetMAP, we use *a priori* information from the simulations (step 6 of Fig 1c) to calculate the discrepancy $\Delta p_j = \hat{p}_j - p_{j,\text{in}}$ between the input and output parameters, where $j$ indexes the parameters. The fractional discrepancy, $\frac{\Delta p_j}{p_{j_\text{in}}} = \frac{\hat{p}_j}{p_{j_\text{in}}} - 1$, has a normal distribution over multiple trials and is shown as box and whisker plots (Fig 2c and Fig. 3e). The percent error for each parameter, the absolute value of the fractional discrepancy:

$$e_j = \frac{|\Delta p_j|}{p_{j_\text{in}}} \cdot 100\%, \quad (S10)$$

describes the accuracy of the output parameters. Because we take an absolute value, $E_j$ has a half-normal distribution. When we repeat the simulated experiment multiple times with different random noise for each repetition, then we obtain a logarithmic average $\bar{E}_j$ over multiple trials of the error. We calculate the average error over parameters as

$$\langle e \rangle = \frac{e_1 + e_2 + \cdots + e_N}{N - D} \quad (S11)$$

where $N$ is the total number of parameters and $D$ is the number of parameters that have been fixed such that their error is zero, which should not contribute to the arithmetic mean. We consider the mean error $\overline{\langle e \rangle}$ averaging over parameters and multiple trials, to be the figure of merit that describes the accuracy of the SVD approach in describing a resonator system, with a lower error $\overline{\langle e \rangle}$ indicating a high accuracy. The input parameters are *a priori* information, only available when conducting simulations, and generally not available when conducting experiments, so we seek to predict the average error $\langle \bar{e} \rangle$ from values available to an experimentalist, including the correlation $R$-value and the signal to noise ratio (SNR). We compute the correlation $R$-value:

$$1 - R^2 = \frac{\text{SSres}}{\text{SStot}} \quad (S12)$$

to quantify the agreement between the input and output spectra (step 7 in Fig 1c). To compute the $R$-value, we solve the equations of motion and calculate the output spectra $\hat{Z}_i(\omega)$ for each resonator $i$ from the recovered parameters $\hat{\vec{p}}$, and plot both the noisy input spectra $\vec{Z}(\omega)$ (colorful datapoints) and the output spectra $\hat{\vec{Z}}(\omega)$ (black dashed line) for $n_R \approx 100$ frequencies, many more frequencies than we use for NetMAP in the examples here. Since we may plot either the amplitude and phase or the real and imaginary parts of the complex amplitude $Z_i$, we have a choice of which plots to use for calculating the $R$-value. For a monomer, we choose both $\text{Re}(Z)$ vs $\omega$ and $\text{Im}(Z)$ vs $\omega$, calculate $R^2$ for each and use the arithmetic mean as our value for $R^2$ in Fig 2d. For a dimer, we choose all four of $\text{Re}(Z_1)$ vs $\omega$, $\text{Re}(Z_2)$ vs $\omega$, $\text{Im}(Z_1)$ vs $\omega$, $\text{Im}(Z_2)$ vs $\omega$, calculate $R^2$ for each and use the arithmetic mean as our value for $R^2$ in Fig 3g. Since we compare parameters output from the SVD analysis and measured spectra, the $R$-values are accessible to experimentalists, whereas directly calculating the error $\overline{\langle e \rangle}$ required *a priori* knowledge of the parameters. By providing a prediction for the error $\overline{\langle e \rangle}$, we will enable an experimentalist to estimate the accuracy of the parameters $\hat{\vec{p}}$ output from the SVD analysis without needing to know *a priori* the true parameters $\vec{p}$.

## 2.6 Factorial methods

Table S1 Factorial parameters for monomer investigation, $2^k r, k = 6, r = 30$

| Factorial for monomer, $2^6$, 29 repeats | − | + |
|---|---|---|
| number of frequencies, $n$ | 2 | 10 |
| Mass $m$ (kg) | 1 | 10 |
| Damping $b$ (N s/m) | 0.1 | 1 |
| Stiffness $k$ (N/m) | 1 | 10 |
| Force amplitude $f$ (N) | 1 | 10 |
| standard deviation $\sigma$ (meters) | $5 \times 10^{-5}$ | $5 \times 10^{-4}$ |

Table S2 Factorial parameters for dimer investigation, $2^k r, k = 8, r = 10$

| Factorial for dimer: $2^8$, 9 repeats | − | + |
|---|---|---|
| mass $m_1$ (kg) | 1 | 10 |
| mass $m_2$ (kg) | 1 | 10 |
| damping $b_1$ (N s/m) | 0.1 | 1 |
| damping $b_2$ (N s/m) | 0.1 | 1 |
| stiffness $k_1$ (N/m) | 1 | 10 |
| stiffness $k_2$ (N/m) | 1 | 10 |
| coupling stiffness $k_{12}$ (N/m) | 1 | 10 |
| Force amplitude $f$ (N) | 1 | 10 |

To understand how the percent error of the algebraic response vector characterization varies with factors of our system, we ran full, replicated two-level factorial experiments—e.g. $2^k$ experiments, [6] where $k = N$. The input factors of these experiments include spring constants, damping, mass, the applied force, the standard deviation of the input noise of simulated spectra, and the number of frequencies sampled. We varied these factors over an order of magnitude. For the monomer study, $k = 6$ parameters, and for the dimer study, $k = 8$. Factors and levels for monomers and dimers are shown in Table S1 and Table S2, respectively. The response variables for this experiment included the average error for one-, two-, and three-dimensional null spaces. In the case of two- and three-dimensional null spaces, we scaled the singular vectors as described above (Methods section 5). We replicated monomer experiments 29 times, for a total of 30 full simulated experiments. We replicated dimer experiments 9 times, for a total of 10 full simulated experiments. The results of the simulations were analyzed using ANOVA factor screening methods, and the model assumptions were checked and verified for each. We analyzed the logarithm of the error $\overline{\langle e \rangle}$ because the raw data did not satisfy the equal variances assumption, and the logarithm of a half-normal distribution better approximates a normal distribution. The logarithm stabilized the variance appropriately. We used $\alpha = 0.05$ significance cutoff for effects and interactions in the reduced model.

## 2.7 Simplifying assumptions

There are several simplifying assumptions we make for this investigation regarding how measurements are made and what information is available to an experimenter. We assume as an approximation that the resonant system is not over-driven, heated, or subject to temperature fluctuations. We assume that masses, spring stiffnesses, and damping coefficients are constant, regardless of the driving frequency or force.

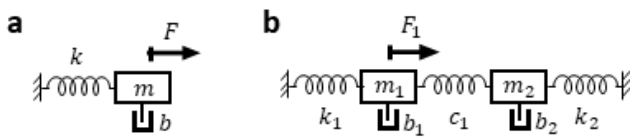

Figure S2 Cartoon of a) monomer and b) dimer system with two coupled masses.

We assume that the experimenter has a correct model for the number of resonant masses and springs and their connections, and that the system has reached a steady-state oscillation, where the energy dissipated by the damping balances the energy input by the force. For this investigation, we consider the simplest two systems (Figure S2): a single damped driven resonant mass on a spring, which we call a monomer, and a resonant dimer system consisting of a pair of two masses, each connected to a wall by a spring and coupled together with a third spring.

For both the monomer and dimer presented here, we assume the experimenter drives one mass with a sinusoidal force $F = f \cos(\omega t)$ and measures the amplitude $A$ and phase $\phi$ of each resonant mass, and the phase is measured with respect to the phase of the driving frequency.

The monomer system (Figure S2a) is fully described by the following parameters: mass $m$, spring stiffness $k$, damping coefficient $b$, and driving force $F(t)$. We assume that the driving force amplitude $f$ and frequency $\omega$ are known. If, however, an experimenter does not know the driving amplitude $f$ in Newtons, knowledge of any one of the other parameters, $m$, $k$, or $b$, is equivalently useful, or the experimenter can use NetMAP to obtain the force-normalized parameters, $m/f$, $k/f$, and $b/f$ without requiring any additional information for a 1D solution space. In our related work, [2] we scale the parameters with respect to the spring constant, calculating $m/k$, $b/k$, and $f/k$. Higher dimensional solution spaces require more *a priori* information about the parameters.

The dimer system (Figure S2b) is fully described by the following parameters: masses $m_1$ and $m_2$, with damping coefficients $b_1$ and $b_2$, each attached to a wall by a spring of stiffnesses $k_1$ and $k_2$ and coupled to each other by a spring of stiffness $c_1$, also called $k_{12}$. We assume that the first mass of the dimer is driven and that there is no external driving force acting on the second mass, though our algebraic approach can be modified to address an additional driving force.

# 3. Monomers
## 3.1 Simulating a monomer

For a monomer (Figure S2a) driven by a sinusoidal force, the equation of motion in complex form is the differential equation

$$m\ddot{x} = f e^{i\omega t} - kx - b\dot{x} \quad (S13)$$

where $x$ is the complex position of the mass, $\dot{x}$ is its complex velocity, $\ddot{x}$ is its complex acceleration, and the real part of the equation is the physical description. We guess a steady-state solution to this differential equation of the form $x = Z e^{i\omega t}$, where $Z$ is the complex amplitude of the position of the oscillating mass, and, through substitution for $x$ and division by the oscillating factor $e^{i\omega t}$, obtain the time-independent equation of motion

$$(-\omega^2 m + i\omega b + k)Z = f \quad (S14)$$

Separating into the real and imaginary parts, we find the equation of motions requires both

$$(-\omega^2 m + k)\,\text{Re}(Z) - f = 0 \text{ and } -\omega b\,\text{Im}(Z) = 0 \quad (S15)$$

We assume an experimenter would measure the complex amplitude $Z = A e^{i\phi}$ of the single oscillator in response to the applied force $f e^{i\omega t}$.

## 3.2 Recovering monomer parameters using SVD

We next calculate the unknown parameters $m$, $b$, and $k$ from the known values $f$, $\omega$, and $Z$. Normally, the response $Z(\omega)$ is measured at a series of driving frequencies $\omega$ and fit with an iterative least-squares approach that optimizes $1 - R^2$, where $R^2$ is the coefficient of determination. We describe an alternative non-iterative approach using singular value decomposition. We require at least $n = 2$ measurements of the complex amplitude, $Z(\omega_1)$ and $Z(\omega_2)$, and with singular value decomposition (SVD), the method supports any number of measurements above two, as follows. For each measurement $Z(\omega_i)$, we know that both Eq 3a and 3b must hold, and we organize these equations as follows:

$$\mathcal{Z}\vec{p} = \vec{0}$$

$$\begin{bmatrix} -\omega_1^2 \operatorname{Re}(Z(\omega_1)) & -\omega_1 \operatorname{Im}(Z(\omega_1)) & \operatorname{Re}(Z(\omega_1)) & -1 \\ -\omega_1^2 \operatorname{Im}(Z(\omega_1)) & \omega_1 \operatorname{Re}(Z(\omega_1)) & \operatorname{Im}(Z(\omega_1)) & 0 \\ -\omega_2^2 \operatorname{Re}(Z(\omega_2)) & -\omega_2 \operatorname{Im}(Z(\omega_2)) & \operatorname{Re}(Z(\omega_2)) & -1 \\ -\omega_2^2 \operatorname{Im}(Z(\omega_2)) & \omega_2 \operatorname{Re}(Z(\omega_2)) & \operatorname{Im}(Z(\omega_2)) & 0 \end{bmatrix} \begin{bmatrix} m \\ b \\ k \\ f \end{bmatrix} = \begin{bmatrix} 0 \\ 0 \\ 0 \\ 0 \end{bmatrix} \quad (S16)$$

where the measurement matrix $\mathcal{Z}$ is calculated from the $n$ spectrum measurements. For the monomer, each frequency contributes two rows to $\mathcal{Z}$, so $n = 2$ frequencies provide a $4 \times 4$ square matrix. We have flexibility in the number of measured frequencies by expanding to a rectangular matrix, where each additional frequency adds an additional two rows to the matrix. We assume an experimenter's measurement of each element of the measurement matrix $\mathcal{Z}$ is subject to Gaussian noise affecting $Z$ but that the noise in frequency $\omega_i$ is negligible. In order to algebraically solve the differential equation, we apply singular value decomposition (SVD) to the matrix to obtain its singular values and corresponding 4-dimensional singular vectors. For a square normal matrix with a normal eigenbasis, we may also call these the eigenvalues and eigenvectors. The singular value corresponding to zero has a singular vector corresponding to the parameters vector $\vec{p}$. We use the known force amplitude $f$ to normalize the parameters vector, thereby obtaining the physical parameters $\hat{m}$, $\hat{b}$, and $\hat{k}$, where the hat symbol indicates that these are calculated from our model of the measured data.

### 3.3   Details for the monomer case shown in Figure 2 of the main text

1. Scaling

Here we provide details about the 1D and 2D scaling for the monomer shown in Fig 2 of the main text. This monomer has:

MONOMER

We set the input values to m = 4 kg, b = 0.01 N s/m, k = 16 N/m, and f = 1 N, so $\omega_{\text{res}} \approx 1.99999922$ rad/s and we approximate the quality factor as $Q \sim \sqrt{\frac{mk}{b}} = 800$. We set the input noise to have a standard deviation of $\sigma = 0.005$ m. We select 2 frequencies for SVD analysis, namely $\omega_a = 1.99999922$ rad/s and $\omega_b = 2.00125039$ rad/s. These are circled in Fig 2ab.

The matrix for SVD analysis is

$\mathcal{Z} = $ [[-1.2629* 10^-1    1.00004* 10^2    3.1573* 10^-2    -1]
        [ 2.00008* 10^2    6.3147* 10^2    -5.0002* 10^1    0]
        [ 1.0007* 10^2    5.0011* 10^1    -2.4985* 10^1    -1]
        [ 1.0009* 10^2    -5.0002* 10^1    -2.4990* 10^1    0]].

The smallest singular value, $\lambda_1 = 1.0676 \times 10^{-7}$, corresponds to singular vector

$$\hat{\vec{p}} = (\hat{m}, \hat{b}, \hat{k}, f) = \alpha\,(-0.2421 \text{ kg}, -0.0006053 \text{ N/(m/s)}, -0.96836 \text{ N/m}, -0.060536 \text{ N}),$$

where

$$\alpha = F_{\text{in}}/-0.060536 = 1/-0.060536 = -16.519$$

is a scaling coefficient obtained from our knowledge of the force amplitude $f$ for a 1D-SVD analysis. Dividing by $\alpha$ allows us to scale the singular vector to yield the modeled parameters vector. We thus obtain $\hat{m} = 3.99912$ kg, $\hat{b} = 0.0099996 \frac{N}{m/s}$, and $\hat{k} = 15.99647$ N/m. The percent errors for each of these is $-0.022\%$, $-0.0040\%$, and $0.022\%$, respectively. Each of these is within 0.022% of the correct values for m, b, and k. We also see that the recovered value

$$\sqrt{\frac{\hat{k}}{\hat{m}}} = 2.000000008 \text{ rad/s}$$

is more accurate than the individually recovered values for mass and spring stiffness. The percent error for $\sqrt{\frac{\hat{k}}{\hat{m}}}$ compared to $\sqrt{\frac{k_{in}}{m_{in}}}$ is $4.1 \times 10^{-7}\%$. This high accuracy likely arises because we choose frequency $\omega_a$ at the peak amplitude and the lightly damped monomer has a sharply peaked resonance, so $\sqrt{k/m}$ is well defined.

2. 2D Solution space for the monomer example

In our example, the vector

$\beta \vec{p}_{2,un} + \alpha \vec{p}_{1,un} = \beta$[-1.62549828e-02, 9.98136999e-03, -5.83247471e-02, 9.98115410e-01] +

$\alpha$ [-2.42090968e-01, -6.05227330e-04, -9.68363873e-01, -6.05227821e-02]

spans the 2D solution space, but we need to determine both $\alpha$ and $\beta$ to obtain the physical parameters. For example, if we assume the experimenter independently knows both $f_{in} = 1$ N and $m_{in} = 4$ kg, then it is straightforward to obtain the coefficients $\alpha = -16.5227$ and $\beta = -6.60069e-7$ to then obtain $\hat{b} = 0.009999992$ and $\hat{k}$ = 15.999999993. Generating the curves from these two recovered parameters yields $R_A^2 = 1 - 1 \times 10^{-11}$ and $R_\phi^2 = 1 - 4 \times 10^{-7}$. In this case, the 1D solution and 2D solution were almost the same. We note that $\alpha \gg \beta$, thus selecting a 2D solution almost entirely in the 1D solution space. The 1D solution is preferred because it requires independent knowledge of only 1 parameter, and we find that it is often reasonably accurate, even when the 2D or 3D solution is more accurate.

3.4 Details for the monomer case shown in Figure 4 of the main text

For the monomer in Figure 4, the maximum number of frequencies is $n = 25$. We set the input values to: $m = 4$ kg, $b = 0.4 \frac{N}{m/s}$, $k = 10$ N/m, $f = 1$ N. Then the resonance frequency $\omega_{res} = 1.58$ rad/s and

$$Q = \sqrt{\frac{mk}{b}} = 15.8.$$

We set the input noise $\sigma = 0.0005$ m. We use up to 25 frequencies for SVD analysis, namely
[1.5795569 1.5961569 1.5629569 1.6127569 1.5463569 1.6293569 1.5297569 1.6459569
1.5131569 1.6625569 1.4965569 1.6791569 1.4799569 1.6957569 1.4633569 1.7123569
1.4467569 1.7289569 1.4301569 1.7455569 1.4135569 1.7621569 1.3969569 1.7787569
1.3803569] rad/s.

The spectrum with these 25 frequencies identified is shown in Figure S3, left. For $n = 25$ frequencies, the $\mathcal{Z}$ matrix has 50 rows. Its smallest singular value, $\lambda_1 = 0.00059708$, corresponds to singular vector $\hat{\vec{p}} = (\hat{m}, \hat{b}, \hat{k}, \hat{f}) = \alpha(-0.36955$ kg, $-0.036957$ N/(m/s), $-0.92387$ N/m, $-0.092383$ N), where $\alpha = f_{in}/-0.092383$ N $= 1/-0.092383 = -10.825$ is a normalization constant obtained from our knowledge of the force amplitude f for a 1D-SVD analysis. Dividing by $\alpha$ allows us to scale the singular vector to yield the modeled parameters vector. Therefore, we obtain $\hat{m} = 4.00017$ kg, $\hat{b} = 0.400043$ N/(m/s) and $\hat{k}$ = 10.00047 N/m. The percent errors for each of these is 0.0041835%, 0.010689%, and 0.0046745%, respectively. Each of these is within 0.010689% of the correct values for m, b, and k. We also see that the recovered value

$$\sqrt{\frac{\hat{k}}{\hat{m}}} = 1.5811 \text{ rad/s}$$

is more accurate than the individually recovered values for mass and spring stiffness; this is often true for various sharply peaked systems. The percent error for $\sqrt{\frac{\hat{k}}{\hat{m}}}$ compared to $\sqrt{\frac{k_{in}}{m_{in}}}$ is 0.00025%. Heatmaps in Figure 4c of the main text show how accurately $n = 2$, 1D-SVD and 2D-SVD recover the solution for this monomer. Figure S3, right shows a similar heatmap for $n = 2$, 3D-SVD. For a monomer, 3D-SVD only recovers damping $b$; all other parameters must be known *a priori* in order to find the solution in the 3D solution space, so the average error is showing the error in damping $\bar{e}_b$ (Figure S3, right).

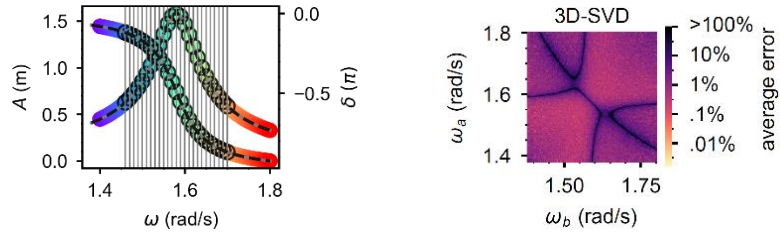

Figure S3. (left) The spectrum also shown in Fig 4a, with corresponding color scale. Vertical lines and circles show the $n = 25$ frequencies used for analysis. $\delta$ is the phase. (right) Heatmap showing the average error for $D = 3$, $n = 2$ analysis of the same monomer system analyzed in Figure 4.

Figure S4 shows how the average error for the monomer parameters decreases for 1D-SVD as the number of frequency points increases. A connecting letters report (Table S3) shows that there are diminishing returns to increasing the number of frequency points: $n = 25$ is not significantly better than $n = 7$. Figure S5 expands on Fig. 4f, showing the variation of the error for 1D-SVD for both of the smallest two singular values. Since the error varies more with $\lambda_2$ than with $\lambda_1$, Figure 4f in the main text shows the variation only with $\lambda_2$.

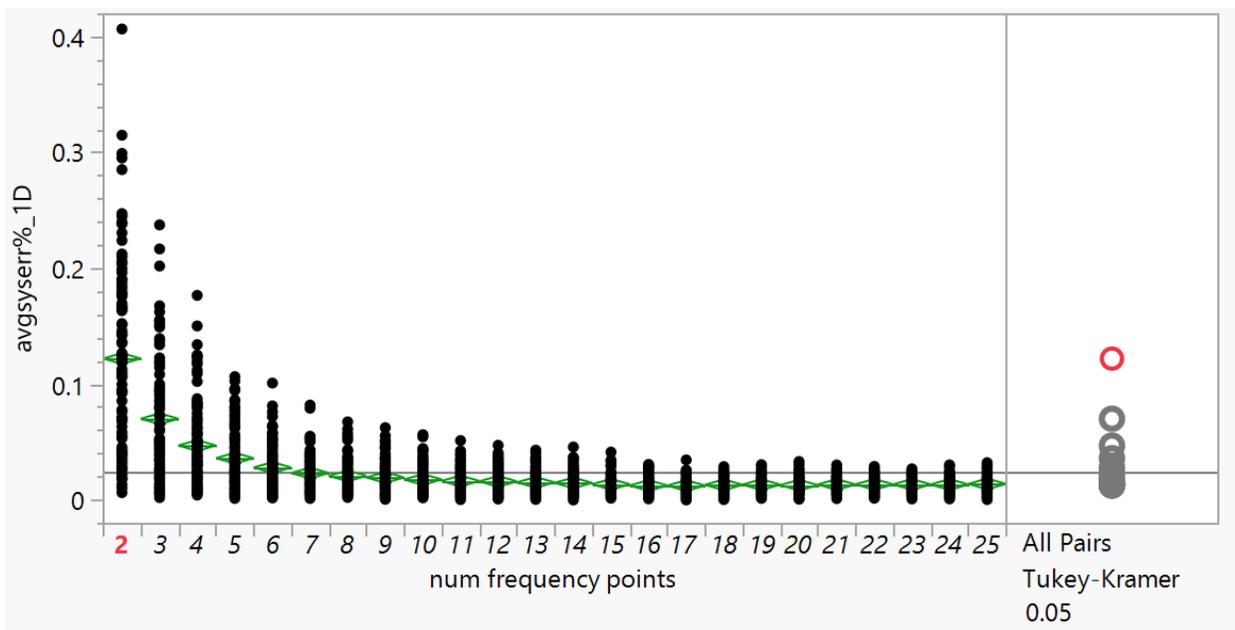

Figure S4. Average error $\langle e \rangle$ for $m, k, b$ as a function of the number of frequency points used in the SVD analysis of a monomer. These are the same datapoints as Fig 4b for 1D-SVD in the main text.

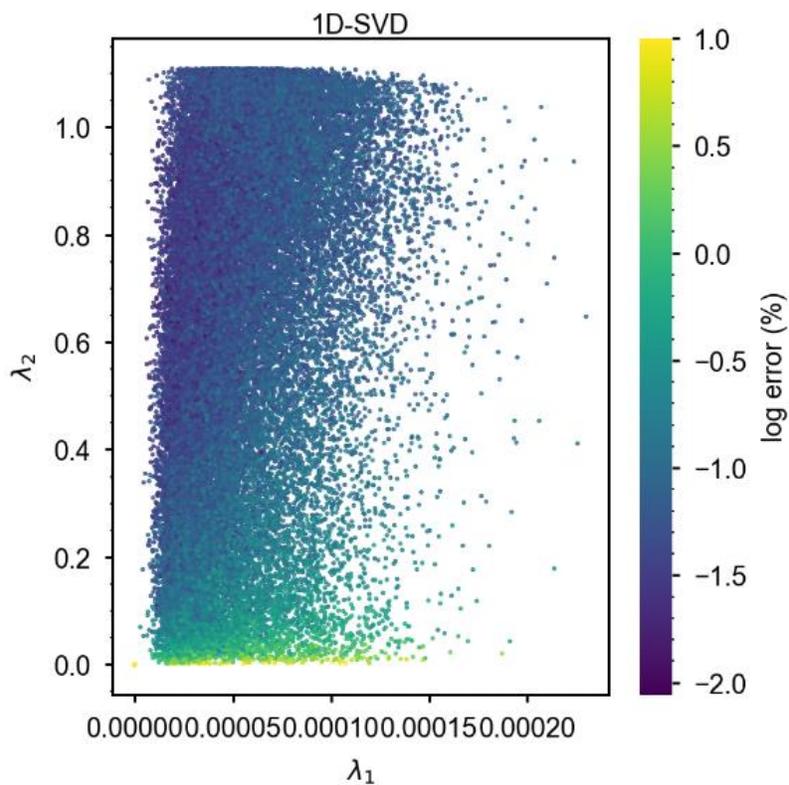

Figure S5. The 1D solution is much less accurate when $\lambda_2$ is small. For $\lambda_2$ near zero, the error of the 1D solution rises to 10%, and for larger $\lambda_2$, the 1D solution error is as low as 0.01%. This is the same dataset shown in Fig, 4f of the main text.

Table S3 Connecting letters report for the data shown in Figure S4, corresponding to Fig. 4b.

**Connecting Letters Report**

| Level | | | | | | Mean |
|---|---|---|---|---|---|---|
| 2 | A | | | | | 0.12234474 |
| 3 | | B | | | | 0.07036778 |
| 4 | | | C | | | 0.04724179 |
| 5 | | | C | D | | 0.03634654 |
| 6 | | | | D | E | 0.02822636 |
| 7 | | | | | E F | 0.02346695 |
| 8 | | | | | E F | 0.02102705 |
| 9 | | | | | E F | 0.01964277 |
| 10 | | | | | E F | 0.01791940 |
| 11 | | | | | E F | 0.01659641 |
| 12 | | | | | E F | 0.01627824 |
| 13 | | | | | F | 0.01542926 |
| 14 | | | | | F | 0.01497419 |
| 25 | | | | | F | 0.01406188 |
| 15 | | | | | F | 0.01372911 |
| 22 | | | | | F | 0.01367425 |
| 23 | | | | | F | 0.01360943 |
| 24 | | | | | F | 0.01357419 |
| 21 | | | | | F | 0.01355960 |
| 19 | | | | | F | 0.01344894 |
| 18 | | | | | F | 0.01328959 |
| 20 | | | | | F | 0.01297662 |
| 16 | | | | | F | 0.01272914 |
| 17 | | | | | F | 0.01266374 |

Levels not connected by same letter are significantly different.

# 4. Dimers
## 4.1 Simulating the dimer

We consider how the algebraic approach presented here applies to a dimer system of damped driven oscillators: two masses $m_1$ and $m_2$, with damping $b_1$ and $b_2$, respectively, each coupled to a wall by a spring of stiffness $k_1$ and $k_2$, respectively, and coupled to each other with a spring of stiffness $k_{12}$ (Figure S2b). The equations of motion of the damped driven dimer in complex form are the coupled differential equations

$$m_1 \ddot{x}_1 = f_1 e^{i\omega t} - b_1 \dot{x}_1 - k_1 x_1 - k_{12}(x_1 - x_2) \text{ and}$$

$$m_2 \ddot{x}_2 = f_2 e^{i\omega t} - b_2 \dot{x}_2 - k_2 x_2 - k_{12}(x_2 - x_1), \tag{S17}$$

where $f_1$ is the amplitude of the driving force applied to the first oscillator and $f_2$ is the amplitude of the oscillating driving force applied to the second oscillator. The two forces are assumed to have the same frequency and phase, such that the force amplitudes $f_1$ and $f_2$ are each real constants, or one of the two is assumed to be zero. We guess a steady-state solution of the form $x_1 = Z_1 e^{i\omega t}$, $x_2 = Z_2 e^{i\omega t}$, where $Z_1 = A_1 e^{i\phi_1}$ and $Z_2 = A_2 e^{i\phi_2}$ are the complex amplitudes of the respective oscillating masses. Then the equations of motion (B1) give us

$$(-m_1 \omega^2 + k_1 + k_{12} + i\omega b_1)Z_1 - k_{12}Z_2 = f_1 \text{ and}$$

$$-k_{12}Z_1 + (-m_2 \omega^2 + k_{12} + k_2 + i\omega b_2)Z_2 = f_2. \tag{S18}$$

In order to simulate the spectra, we organize these two equations into a matrix equation,

$$\begin{bmatrix} -m_1 \omega^2 + k_1 + k_{12} + i\omega b_1 & -k_{12} \\ -k_{12} & -m_2 \omega^2 + k_{12} + k_2 + i\omega b_2 \end{bmatrix} \begin{bmatrix} Z_1 \\ Z_2 \end{bmatrix} = \begin{bmatrix} f_1 \\ f_2 \end{bmatrix} \tag{S19}$$

and use Cramer's rule to calculate the spectra $z_1(\omega)$ and $z_2(\omega)$ for a given set of physical parameters $\vec{p}_{\text{in}} = (m_1, m_2, b_1, b_2, k_1, k_2, k_{12}, f_1, f_2)$. As before, we add random Gaussian noise $\Gamma_{x,i}(\sigma, \omega) + i\Gamma_{y,i}(\sigma, \omega)$ to the spectra to simulate noisy experimental conditions.

## 4.2 Recovering the dimer parameters with SVD

Having now simulated an experiment, we consider how we can recover the physical parameters from the noisy spectra $Z_1(\omega)$ and $Z_2(\omega)$ and knowledge of the driving forces $F_1 = f_1 e^{i\omega t}$ and $F_2 = f_2 e^{i\omega t}$. We reorganize Eq (B2) into a different matrix equation than we used for Cramer's rule by creating a vector $\vec{p}$ of parameters and a matrix $\mathcal{Z}$ of measurements such that:

$$\mathcal{Z} \vec{p} = \vec{0}$$

$$\begin{bmatrix}
-\omega_1^2 X_{11} & 0 & -\omega_1 Y_{11} & 0 & X_{11} & 0 & X_{11}-X_{21} & -1 & 0 \\
-\omega_1^2 Y_{11} & 0 & \omega_1 X_{11} & 0 & Y_{11} & 0 & Y_{11}-Y_{21} & 0 & 0 \\
0 & -\omega_1^2 X_{21} & 0 & -\omega_1 Y_{21} & 0 & X_{21} & X_{21}-X_{11} & 0 & -1 \\
0 & -\omega_1^2 Y_{21} & 0 & \omega_1 X_{21} & 0 & Y_{21} & Y_{21}-Y_{11} & 0 & 0 \\
\vdots & \vdots & \vdots & \vdots & \vdots & \vdots & \vdots & \vdots & \vdots \\
-\omega_n^2 X_{1n} & 0 & -\omega_n Y_{1n} & 0 & X_{1n} & 0 & X_{1n}-X_{2n} & -1 & 0 \\
-\omega_n Y_{1n} & 0 & \omega_n X_{1n} & 0 & Y_{1n} & 0 & Y_{1n}-Y_{2n} & 0 & 0 \\
0 & -\omega_n^2 X_{2n} & 0 & -\omega_n Y_{2n} & 0 & X_{2n} & X_{2n}-X_{1n} & 0 & -1 \\
0 & -\omega_n^2 Y_{2n} & 0 & \omega_n X_{2n} & 0 & Y_{2n} & Y_{2n}-Y_{1n} & 0 & 0
\end{bmatrix}\begin{bmatrix} m_1 \\ m_2 \\ b_1 \\ b_2 \\ k_1 \\ k_2 \\ k_{12} \\ f_1 \\ f_2 \end{bmatrix} = \begin{bmatrix} 0 \\ 0 \\ 0 \\ 0 \\ 0 \\ 0 \\ 0 \\ 0 \\ 0 \end{bmatrix}$$

where $X_{ji} = \text{Re}\left(Z_j(\omega_i)\right)$ and $Y_{ji} = \text{Im}\left(Z_j(\omega_i)\right)$ provide information about measurements of each resonator at each of the discrete frequency points.

The measurement matrix $\mathcal{Z}$ has $\dim(\vec{p})$=9 columns and $4n$ rows for the dimer, where $n$ is the number of frequency points used for the analysis. In general we have $N_{\text{cluster}}$ complex equations describing the system. Taking the real and imaginary parts, we rewrite these as $2N_{\text{cluster}}$ real equations and thus the real measurement matrix $\mathcal{Z}$ has $2nN_{\text{cluster}}$ rows. For example, using measurements from two frequencies to describe a dimer would require an $8 \times 9$ matrix whereas using 30 frequencies would require a $120 \times 9$ matrix $\mathcal{Z}$. We then analyze the measurement matrix $\mathcal{Z}$ using SVD to identify the singular vectors and identify 1 or more singular vectors corresponding to the singular value of zero. The dimension of the solution space may be open to interpretation because the measurements are noisy, so the smallest singular values do not precisely equal zero. If we interpret the solution space to be 1-dimensional then there is one singular vector corresponding to the null singular value. We therefore scale the singular vector using our knowledge of one element of the parameters vector $\vec{p}$. In particular, here we assume we know force amplitude $f_1$ pushing the first mass. Hence the analysis recovers all but one of the elements of the parameters vector $\vec{p} = (m_1, m_2, b_1, b_2, k_1, k_2, k_{12}, f_1, f_2)$. The resulting output parameters $\hat{p}_i$ are not exactly equal to the input parameters $p_{i,\text{in}}$ due to noise in the measurements. If we do not add noise, the recovered parameters are exactly equal to the input parameters, within computer precision. Therefore, we add noise in order to validate NetMAP with noise.

### 4.3 Statistical comparison tests for dimer case
Table S4 shows the statistical comparison tests for the dimer case shown in Figure 3 of the main text.

Table S4. Observed $t$-statistics and associated $p$-values for network parameters of the dimer shown in Fig. 3 of the main text.

| Null-space Dimension | $\hat{p}_j$ | $t_0 \equiv \dfrac{\lvert (p_{\text{in}})_j - \hat{p}_j \rvert}{s.e._j}$ | $p$-value |
|---|---|---|---|
| 1D | $\hat{m}_1$ | 1.8867 | 0.0595 |
| 1D | $\hat{b}_1$ | 0.3242 | 0.7458 |
| 1D | $\hat{k}_1$ | 1.7826 | 0.0750 |
| 1D | $\hat{m}_2$ | 1.2347 | 0.2172 |
| 1D | $\hat{b}_2$ | 0.4928 | 0.6223 |
| 1D | $\hat{k}_2$ | 1.2431 | 0.2141 |
| 1D | $\hat{k}_{12}$ | 1.9109 | 0.0563 |
| 2D | $\hat{b}_1$ | 0.9078 | 0.3642 |
| 2D | $\hat{k}_1$ | 0.9317 | 0.0750 |
| 2D | $\hat{m}_2$ | 1.1388 | 0.2551 |
| 2D | $\hat{b}_2$ | 0.7423 | 0.4581 |
| 2D | $\hat{k}_2$ | 1.1114 | 0.2667 |
| 2D | $\hat{k}_{12}$ | 0.1713 | 0.8640 |

## 4.4 Error is inversely proportional to the SNR

Since the error is proportional to the input noise $\sigma$, the error is also inversely proportional to the SNR, as shown in Figure S6.

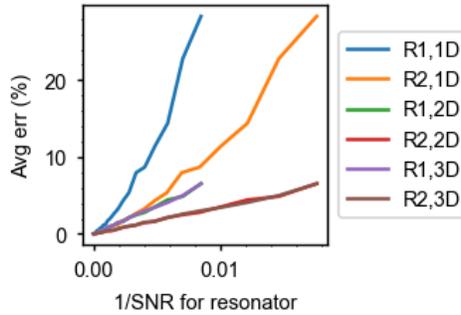

Figure S6 For the dimer shown in Fig. 3 of the main text, error is inversely proportional to the SNR. This is another view of the data shown in Fig. 3h of the main text, where input noise is varied. The average SNR for resonator 1 (R1) and resonator 2 (R2) also varies, and the error of the recovered parameters is proportional to the input noise. The lines here are noisy but linear.

## 4.5 Details for the dimer case shown in Figure 5 of the main text

Here we provide details relating to the dimer case shown in Figure 5 of the main text. Figure S7 shows the R1 and R2 spectra for this dimer. The second resonance peak in (a), near 3.501 rad/s, is imperceptible on the amplitude spectrum but does show an effect in the phase spectrum. Figure 5c of the main text shows how the accuracy of the 1D solution improves as additional response vectors are incorporated into the analysis. The improvement shows diminishing benefits as the number of response vectors becomes large. A connecting letters report (see Table S5) shows which of the errors are significantly different from each other. Levels sharing a letter are not significantly different from each other.

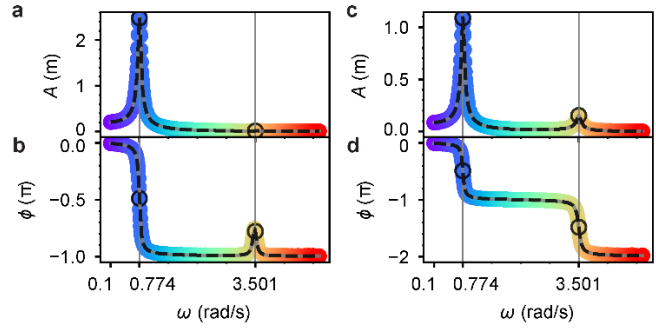

Figure S7. Spectra corresponding to Fig. 5 of the main text. **a)** Amplitude $A_1(\omega)$, **b)** Phase $\phi_1(\omega)$, **c)** Amplitude $A_2(\omega)$, **d)** Phase $\phi_2(\omega)$. The black circles and vertical grey lines indicate the two input frequencies $\omega_a = 0.774$ rad/s and $\omega_b = 3.501$ rad/s, each at a resonant peak, used for the SVD analysis ($n = 2$. The subtle grey curves show the exact spectra calculated from input parameters using Cramer's rule. The colorful datapoints show simulated spectra with added noise $\sigma = 5 \times 10^{-5}$ m. The dashed black line shows the output spectra $\hat{A}$ and $\hat{\phi}$ calculated from the 1D-SVD recovered parameters $\hat{\vec{p}}$ for one trial, showing agreement with the exact spectra in grey and with the noisy simulated spectra in rainbow colors.

Table S5 Connecting letters report for the 1D solutions shown in Fig 5c of the main text.

| n | | Avg err (%) |
|---|---|---|
| 2  | A                     | 0.50053195 |
| 3  | A                     | 0.48953156 |
| 4  | B                     | 0.38672077 |
| 5  | B C                   | 0.37369670 |
| 6  | B C D                 | 0.31920998 |
| 7  | C D E                 | 0.31322216 |
| 8  | D E F                 | 0.27450738 |
| 9  | D E F G               | 0.26417585 |
| 10 | E F G H               | 0.24553969 |
| 11 | F G H I               | 0.24172491 |
| 12 | F G H I J             | 0.22021185 |
| 13 | F G H I J K           | 0.21479855 |
| 14 | G H I J K L           | 0.19807178 |
| 15 | H I J K L M           | 0.19437583 |
| 16 | H I J K L M N         | 0.19214605 |
| 17 | H I J K L M N         | 0.18880677 |
| 18 | H I J K L M N         | 0.18361781 |
| 19 | H I J K L M N         | 0.18023757 |
| 20 | I J K L M N           | 0.17378932 |
| 21 | J K L M N             | 0.17000446 |
| 22 | J K L M N             | 0.16381124 |
| 23 | J K L M N             | 0.16225726 |
| 24 | J K L M N             | 0.15636442 |
| 25 | J K L M N             | 0.15440887 |
| 26 | J K L M N             | 0.15140726 |
| 27 | K L M N               | 0.15041748 |
| 28 | K L M N               | 0.14757924 |
| 29 | K L M N               | 0.14580769 |
| 30 | L M N                 | 0.14003248 |
| 31 | L M N                 | 0.13892956 |
| 32 | L M N                 | 0.13846269 |
| 33 | L M N                 | 0.13749258 |
| 34 | L M N                 | 0.13643984 |
| 35 | L M N                 | 0.13580993 |
| 36 | L M N                 | 0.13500138 |
| 37 | L M N                 | 0.13462160 |
| 42 | L M N                 | 0.13281584 |
| 43 | L M N                 | 0.13217349 |
| 40 | L M N                 | 0.13186677 |
| 44 | L M N                 | 0.13148252 |
| 41 | L M N                 | 0.13144607 |
| 38 | L M N                 | 0.13142821 |
| 45 | L M N                 | 0.13102673 |
| 39 | L M N                 | 0.13085457 |
| 46 | L M N                 | 0.12903261 |
| 47 | L M N                 | 0.12863958 |
| 48 | M N                   | 0.12781030 |
| 49 | M N                   | 0.12708705 |
| 50 | N                     | 0.12470819 |

# 5. Experimental considerations for measuring absolute phase

Measuring the absolute phase requires some experimental consideration because the time elapsing between the driving force and the response usually is measured by a lock-in amplifier, and the electronics and optics will introduce delays before the driving force arrives at the resonator system and additional delays before the measurement can be read, creating a phase delay $\theta = \omega t_{\text{delay}}$, which is added onto the absolute phase $\phi$. We assume that this time delay can be measured and subtracted to obtain an absolute phase, as follows [2]. An experimenter may obtain the time delay by measuring the spectra at lower frequencies, where the force and response are expected to be in phase such that the absolute phase $\phi$ is zero. Then the total phase $\theta + \phi = \omega t_{\text{delay}} + 0$, such that a linear fit of the phase versus frequency provides the slope $t_{\text{delay}}$. The phase delay $\theta$ is subtracted for each frequency $\omega$ to obtain the absolute phase $\phi$.

# 6. Additional Cases

## 6.1 Heavily damped monomer

As an additional example, we consider a heavily damped monomer (step 1). We simulate a spectrum for a mass-and-spring where we set $m = 4$ kg, $b = 8$ N/(m/s), $k = 9$ N/m, $f = 1$ N, and Gaussian noise with standard deviation $\sigma = 0.0005$ m (step 2). With these input parameters, the quality factor is $Q \approx \frac{\sqrt{km}}{b} = 0.75$ and the resonant frequency is

$$\omega_r = \sqrt{\frac{k}{m} - \frac{b^2}{2m^2}} = 0.5 \text{ rad/s}.$$

(The approximate equation $\sqrt{\frac{k}{m}} = 1.5$ rad/s greatly overestimates the resonant frequency for the heavily damped case.) We mimic an experiment by simulating the spectrum with noise. We can use any frequency points for the analysis with SVD, and to demonstrate this we choose $n = 3$ frequencies at random, $\omega_1 =$0.183389 rad/s, $\omega_2 =$0.545455 rad/s, and $\omega_2 = $ 1.481318 rad/s, as shown in Figure S8. At these three measurement frequencies, we obtain amplitude $A(\omega_1) =$0.111738 m, $A(\omega_2) = $ 0.112556 m, $A(\omega_3) =$0.083927 m and phase $\phi(\omega_1) =$-0.155537 rad, $\phi(\omega_2) = $ -0.507897 rad, and $\phi(\omega_3) =$-1.542800 rad (step 3). The signal to noise ratio for each measurement is $\frac{A(\omega_1)}{\sigma} =$223, $\frac{A(\omega_2)}{\sigma} = 224$ and $\frac{A(\omega_3)}{\sigma} = 169$, and the mean SNR is 205. We use a parameters vector $\vec{p} = (m, k, b, f)$ and thus construct the real rectangular $6 \times 4$ matrix (step 4)

$$\mathcal{Z} = \begin{bmatrix} -\omega_1^2 \operatorname{Re}(Z(\omega_1)) & -\omega_1 \operatorname{Im}(Z(\omega_1)) & \operatorname{Re}(Z(\omega_1)) & -1 \\ -\omega_1^2 \operatorname{Im}(Z(\omega_1)) & \omega_1 \operatorname{Re}(Z(\omega_1)) & \operatorname{Im}(Z(\omega_1)) & 0 \\ -\omega_2^2 \operatorname{Re}(Z(\omega_2)) & -\omega_2 \operatorname{Im}(Z(\omega_2)) & \operatorname{Re}(Z(\omega_2)) & -1 \\ -\omega_2^2 \operatorname{Im}(Z(\omega_2)) & \omega_2 \operatorname{Re}(Z(\omega_2)) & \operatorname{Im}(Z(\omega_2)) & 0 \\ -\omega_3^2 \operatorname{Re}(Z(\omega_3)) & -\omega_3 \operatorname{Im}(Z(\omega_3)) & \operatorname{Re}(Z(\omega_3)) & -1 \\ -\omega_3^2 \operatorname{Im}(Z(\omega_3)) & \omega_3 \operatorname{Re}(Z(\omega_3)) & \operatorname{Im}(Z(\omega_3)) & 0 \end{bmatrix} \quad (S20)$$

and obtain the singular vector (step 5)

$$\hat{\vec{p}}_{1D} = (\hat{m},\, \hat{b},\, \hat{k},\, f) = \alpha\left(-0.21709 \text{ kg}, -0.50854 \frac{\text{N}}{\text{m/s}}, -0.83167 \frac{\text{N}}{\text{m}}, -0.05080 \text{ N}\right),$$

which we scale with the known input force amplitude $f = 1$ N such that $\alpha = -0.05080$ to obtain

$$(\hat{m},\, \hat{b},\, \hat{k}, f) = \left(4.2734 \text{ kg}, 10.010 \frac{\text{N}}{\text{m/s}}, 16.371 \frac{\text{N}}{\text{m}}, 1\text{N}\right).$$

Comparing $\vec{p}_{in}$ and $\hat{\vec{p}}$, the error for each parameter is $e_m = 6.8\%$, $e_b = 0.10\%$, and $e_k = 2.3\%$, for an average of $\langle e \rangle = 3.1\%$ (step 6).

Calculating the resonance frequency from the recovered parameters is not very accurate when the resonance frequency is so broadened by the heavy damping. Whereas the lightly damped monomer in the main text has high accuracy for recovering $\sqrt{\frac{\hat{k}}{\hat{m}}}$, for this heavily damped monomer, the calculation is less accurate. We have $\sqrt{\frac{\hat{k}}{\hat{m}}} = 1.9573$ rad/s with error $e_{\sqrt{k/m}} = 30\%$, and

$$\sqrt{\frac{\hat{k}}{\hat{m}} - \frac{\hat{b}^2}{2\hat{m}^2}} = 1.0427 \text{ rad/s}$$

with error 109%. These errors for estimated resonance frequency are much higher than the error for the individual parameters, suggesting that we do not have correlation errors for this case.

Hence a hypothetical experimenter would obtain $\hat{m}$, $\hat{b}$, and $\hat{k}$ with an average error $e$ of 3.1% using just three randomly selected measurements of $Z$ with an average signal to noise ratio of 205.

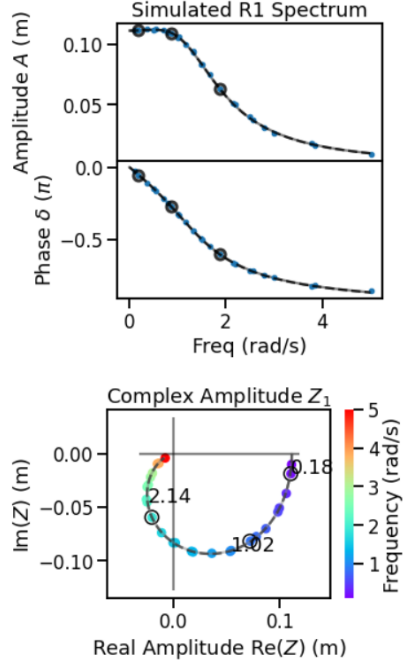

Figure S8 An example of a heavily damped monomer and 1D-SVD results. In each plot, a subtle grey curve shows the simulated spectrum $z(\omega)$ without noise. Colorful datapoints show the simulated noisy spectrum $Z(\omega) = Ae^{i\delta}$, with color indicating frequency. The three circled points in each plot represent the three randomly selected frequencies that are used for analysis. The black dashed curves show the recovered spectrum $\hat{Z}(\omega)$.

## 6.2 Medium-coupled dimer

As an additional example, we consider a dimer with medium coupling, and consider applying an oscillating force to resonator 1 (R1). We set the input values to $m_1 = 11$ kg, $b_1 = 0.5$ N s/m, $k_1 = 9$ N/m, $f_1 = 1$ N, $m_2 = 5$ kg, $b_2 = 0.1$ N s/m, $k_2 = 20$ N/m, and $k_{12} = 4$. The quality factors are

$$Q_1 \approx \sqrt{\frac{m_1 k_1}{b_1}} = 20 \text{ and } Q_2 \approx \sqrt{\frac{m_2 k_2}{b_2}} = 100.$$

We set the input noise to $\sigma = 0.0005$ m. We measure at the two resonance frequencies, 1.05 and 2.21 rad/s. Then the mean SNR for resonator 1 measurements is 193 and for resonator 2 measurements is 83.3. The spectra, histogram, and error versus noise are shown in Figure S9. For this dimer system, we find that the 3D solution is slightly more accurate than the 1D solution, and the 2D solution is significantly less accurate.

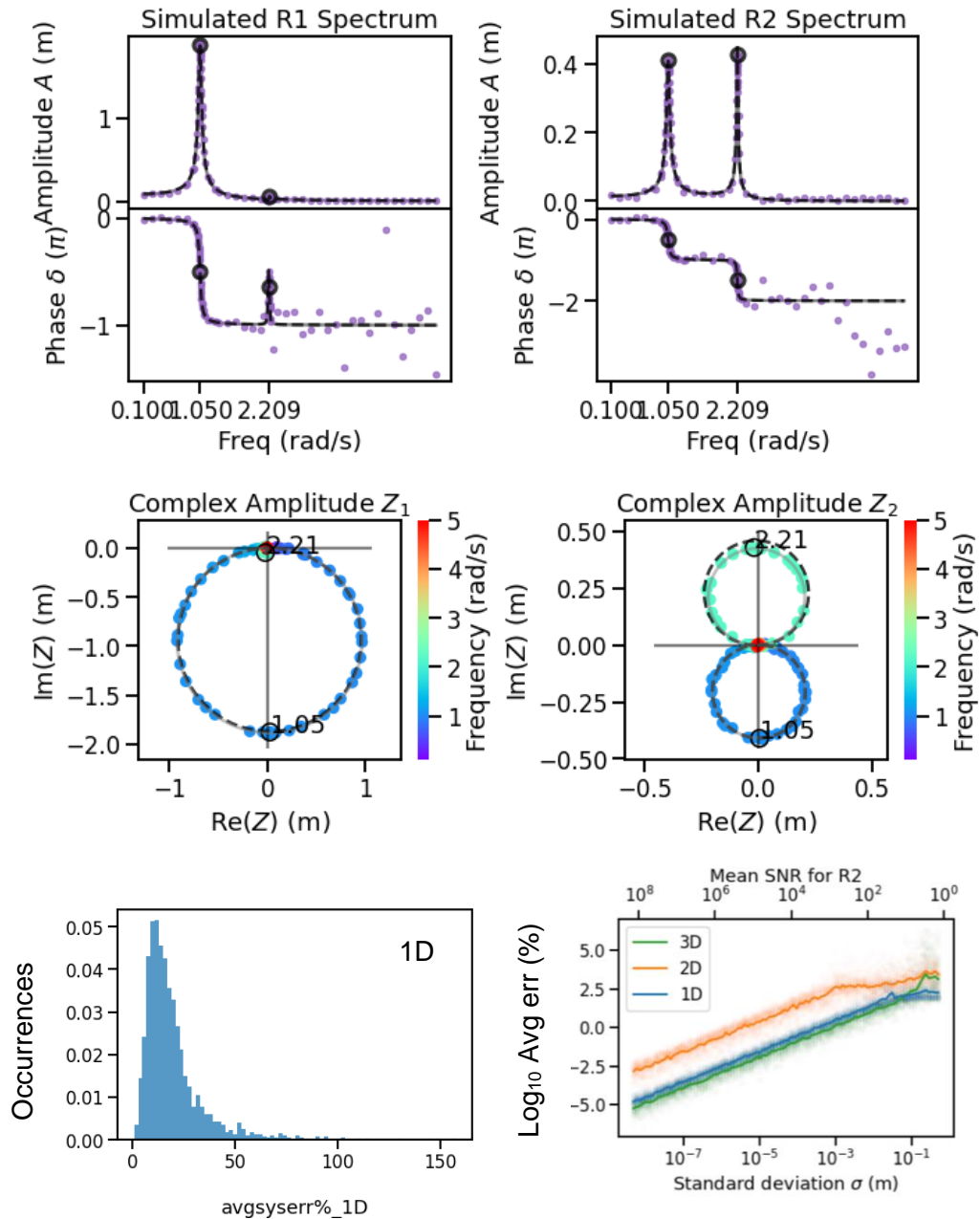

Figure S9. A medium-damped dimer. Resonator 1 has one strong resonance peak and one weak resonance peak, while the two resonance peaks for resonator 2 are equally strong and appear as two loops on the complex plane. The two peak frequencies (circled) are identified for SVD analysis. The 1D-SVD output spectra $\hat{\vec{Z}}(\omega)$ are shown as dashed black lines, and we see that the amplitude of the 2.2 rad/s peak for resonator 2 is not perfectly recovered. The histogram shows the distribution of $\langle e \rangle$ for 1D-SVD.

## 6.3 Strongly coupled dimer

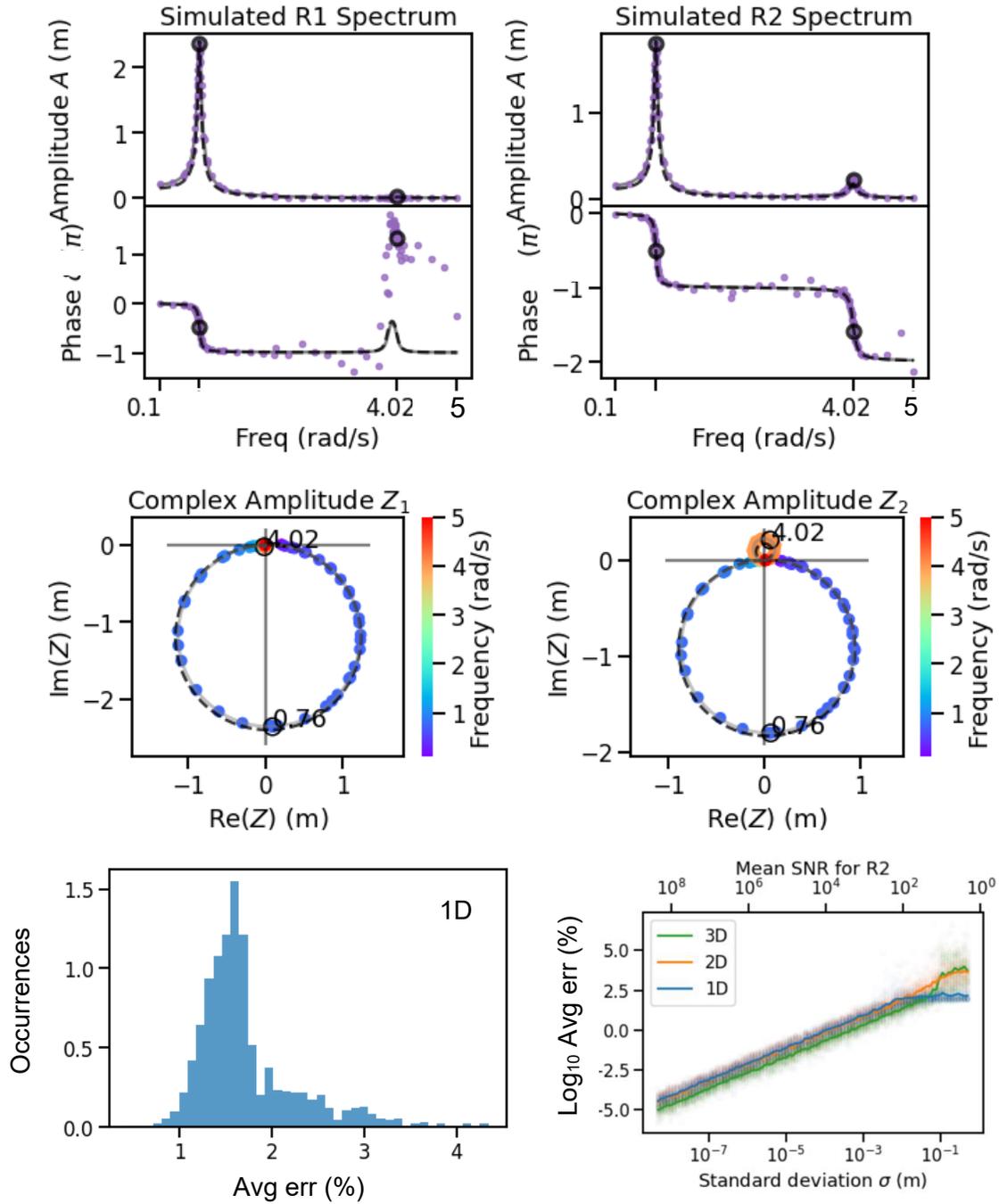

Figure S10. An example with a strongly coupled dimer. The spectra and histogram show a 1D solution.

As an additional example, we consider a dimer with strong coupling, and consider applying an oscillating force to resonator 1 (R1). We set the input values to $m_1 = 8$ kg, $b_1 = 0.5$ N s/m, $k_1 = 2$ N/m, $f_1 = 1$ N, $m_2 = 1$ kg, $b_2 = 0.1$ N s/m, $k_2 = 4$ N/m, and $k_{12} = 11$ N/m. We set the input noise to $\sigma =$

0.0025 m. We measure at the two resonance frequencies, 0.758 and 4.02 rad/s. Then the mean SNR for resonator 1 measurements is 239 and for resonator 2 measurements is 202. For the strongly coupled dimer shown in Figure S10, we find that the 3D solution is the most accurate, while the 1D and 2D solutions are similar to each other.

### 6.4 Force applied to both resonators of a dimer

As an additional example, we consider applying a force $\vec{F} = \langle f \cos \omega t, f \cos \omega t \rangle$ to a dimer. That is, applying the same oscillating force to both masses. We set the input values to: $m_1 = 1$ kg, $b_1 = 0.1$ N s/m, $k_1 = 1$ N/m, $f_1 = f_2 = 10$ N, $m_2 = 10$ kg, $b_2 = 0.1$ N s/m, $k_2 = 10$ N/m, $k_{12} = 1$ N/m, $\sigma = 0.005$ m, and we measure at the resonance frequencies, 1.00 and 1.45 rad/s. Then the mean SNR for resonator 1 measurements is 16130 and for resonator 2 measurements is 10630. The results are shown in Figure S11.

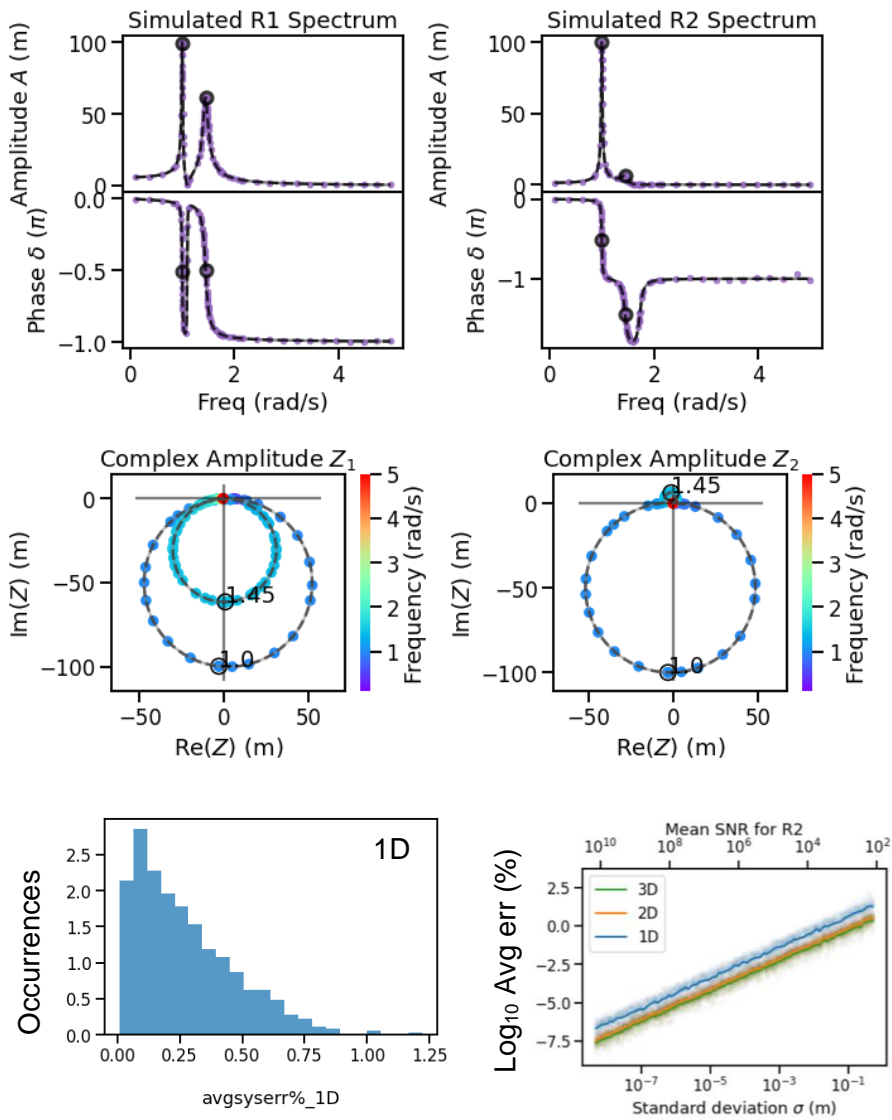

Figure S11 An example of a dimer where the same oscillating force is applied to both resonators

For dimer system in Figure S11 with force applied to both resonators, we find that the 3D solution is slightly more accurate than the 2D solution, and the 1D solution is the least accurate. The amplitude $A_2$ of the second resonator is higher because it is also driven, and therefore the signal to noise ratio for the second resonator is better than cases where only one of the two resonators is driven. This explains why the solutions are overall more accurate than for dimer cases with only one driven oscillator.

# 7. Factorial experiments varying every parameter between two levels

Plots summarizing the factorial experiments are provided here for monomers (Figure S12) and dimers (Figure S13 and Figure S14).

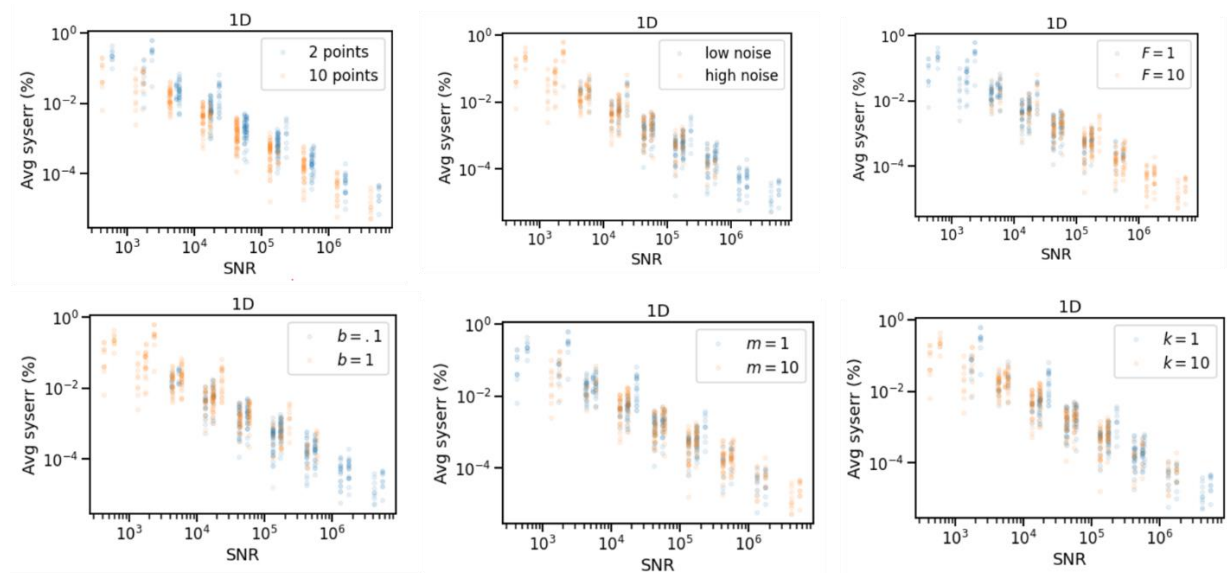

Figure S12 The 1D-SVD error for a monomer is inversely proportional to the SNR, and varies with $n$, $m$, and $b$. Each plot shows the complete set of data from the $2^k r$ experiments for the monomer described in Table S1. The color coding in each varies to show how the relationship between average SNR and average error varies with each of the six varied parameters, $2^k r, k = 6, r = 30$.

An experimentalist using NetMAP will wish to know the accuracy of the parameters vector without knowing the input parameters. A simplified linear model using $R^2$ and mean SNR (Figure S13) can estimate the average error for a wide range of situations.

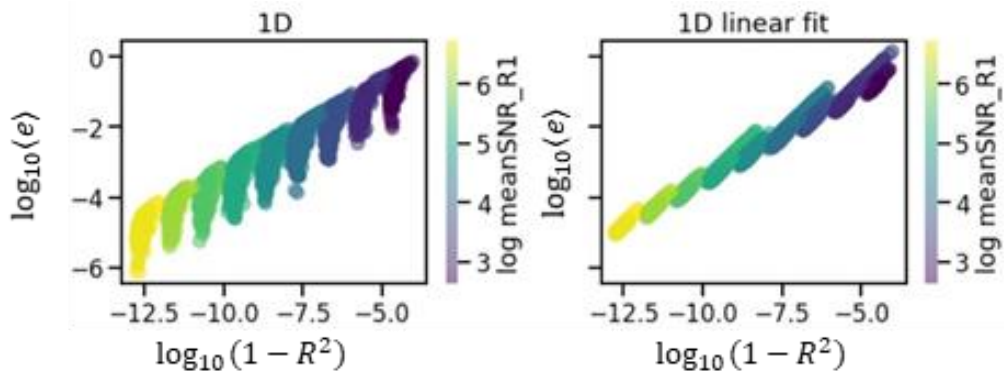

Figure S13. The results of the factorial simulations for dimers are shown on the left, showing that the error $\langle e \rangle$ (vertical axis) varies with $1 - R^2$ (horizontal axis) and SNR (color scale). On the right, a linear fit to the simulated data offers a simplified model for estimating the error from the SNR and $R^2$ values.

| Source | LogWorth | P Value |
|---|---|---|
| f | 187.540 | 0.00000 |
| m1 | 86.153 | 0.00000 |
| m2 | 56.667 | 0.00000 |
| k12 | 48.017 | 0.00000 |
| k1*k2 | 44.900 | 0.00000 |
| m2*k2 | 30.994 | 0.00000 |
| m2*b2 | 27.375 | 0.00000 |
| m2*k12 | 20.579 | 0.00000 |
| b2*k12 | 17.474 | 0.00000 |
| m1*k1 | 14.895 | 0.00000 |
| k2 | 13.679 | 0.00000 |
| m1*k2 | 8.421 | 0.00000 |
| k2*k12 | 7.735 | 0.00000 |
| m2*b1 | 6.947 | 0.00000 |
| m2*k1 | 5.017 | 0.00001 |
| m1*b1 | 4.434 | 0.00004 |
| m1*b2 | 3.765 | 0.00017 |
| b2*k1 | 3.027 | 0.00094 |
| b1 | 2.990 | 0.00102 |
| b1*b2 | 2.129 | 0.00744 |
| m1*k12 | 2.044 | 0.00904 |
| k1 | 1.496 | 0.03194 |
| m1*m2 | 1.415 | 0.03848 |
| k1*k12 | 1.319 | 0.04800 |
| b2 | 0.083 | 0.82623 |

Figure S14 Reduced model effects summary of factorial experiment on the dimer system. The stronger effects and interactions have larger LogWorth values. Plots were generated in JMP software v 16.1.